\documentclass[%
 pra, twocolumn,
 superscriptaddress,
 longbibliography,
 aps,
 showkeys, floatfix]{revtex4-2}

\usepackage[pdftex]{hyperref,color,graphicx}
\usepackage{amsfonts,amssymb,amsmath}
\usepackage[separate-uncertainty=false]{siunitx}
\usepackage[english]{babel}
\usepackage[utf8]{inputenc}
\usepackage[T1]{fontenc}
\usepackage[toc,page]{appendix}

\usepackage{graphicx}
\usepackage[space]{grffile}
\usepackage{placeins}
\usepackage{latexsym}
\usepackage{textcomp}
\usepackage{longtable}
\usepackage{multirow,booktabs}
\usepackage{url}
\hypersetup{colorlinks=false,pdfborder={0 0 0}}
\usepackage[english]{babel}
\usepackage{dcolumn}
\usepackage{bm}
\usepackage{upgreek}
\usepackage{color}
\usepackage{cancel}
\usepackage{hyperref}
\usepackage{braket}

\newcommand{\bbN}{\mathbb{N}}

\begin{document}

\title{Spontaneous-emission induced ratchet in atom-optics kicked rotor quantum walks}

\author{Nikolai Bolik}
\affiliation{Institute of Theoretical Physics, Heidelberg University, Philosophenweg 16, 69120 Heidelberg, Germany}

\author{Sandro Wimberger}
\email{sandromarcel.wimberger@unipr.it}
\affiliation{Dipartimento di Scienze Matematiche, Fisiche e Informatiche, Universit\`{a} di Parma, Parco Area delle Scienze 7/A, 43124 Parma, Italy}
\affiliation{INFN, Sezione di Milano Bicocca, Gruppo Collegato di Parma, Parco Area delle Scienze 7/A, 43124 Parma, Italy}

\begin{abstract}
Quantum walks have gained significant attention over the past decades, mainly because of their variety of implementations and applications. Atomic quantum walks are typically subject to spontaneous emissions arising from the control fields. We investigate spontaneous emission in an atom optics kicked rotor quantum walk. Here, spontaneous emission occurs naturally due to the driving by the kicks, and it is generally viewed as a 
nuisance in the experiment. We find, however, that 
spontaneous emission may induce asymmetries in an otherwise symmetric quantum walk. Our results underscore the utility of spontaneous emission and the application of the asymmetric evolution in the walker's space, i.e. for the construction of a quantum walk ratchet or for Parrondo-like quantum games.
This highlights the potential for reinterpreting seemingly adverse effects as beneficial under certain conditions, thus broadening the scope of quantum walks and their applications.

\end{abstract}

\keywords{Atom-optics kicked rotor, Quantum walks, Bose-Einstein condensates, Floquet-Bloch engineering, Quantum chaos, Quantum-to-classical transition}

\maketitle

\section{'Introduction}
\label{sec-Introduction}

Over the past two decades, quantum walks (QWs) have gained considerable attention owing to their potential to outperform classical algorithms in a range of computational tasks \cite{Kempe2003, Portugal, Kendon}. The distinct characteristics of QWs stem from quantum interference phenomena. For instance, they exhibit a ballistic expansion \cite{Kempe2003, Portugal} as opposed to the diffusive expansion observed in classical random walks, which are governed by stochastic processes \cite{Weiss94}.

Much like their classical counterparts, QWs can be broadly divided into two primary types: The discrete-time and the continuous-time QWs \cite{Kempe2003}. Notably, the discrete-time variant introduces an additional coin-degree-of-freedom, which dictates the direction of the walker in the subsequent step.

In preceding studies, we have investigated a distinctive model, describing discrete-time QW, employing a Bose-Einstein condensate (BEC), consisting of $^{87}$Rb atoms, with an additional spin-1/2 degree of freedom \cite{Romanelli2013, Clark2021,dadras2019experimental, Bolik2022, Dadras2018}. This approach markedly differs from various other experimental investigations \cite{preiss2015strongly,dur2002quantum, eckert2005one, steffen2012digital, groh2016robustness,karski2009quantum, chandrashekar2006implementing, travaglione2002implementing, zahringer2010realization, schmitz2009quantum, perets2008realization, peruzzo2010quantum, cardano2017detection, chen2018observation, tang2018experimental, poulios2014quantum, schreiber2010photons}, distinguishing itself by realizing 
the QW to occur within quantized momentum space, where the steps of the walk are implemented by periodic pulses applied to the condensate. This process is meticulously described by the atom-optics kicked rotor (AOKR) model \cite{Raizen1999, SW2011, Rev2022}, governed by the following Hamiltonian:
\begin{equation}
\hat {\mathcal{H}} = \frac{\hat p^2}{2} + k\cos(\hat \theta)\sum_{j=-\infty}^{\infty}\delta(t-j\tau).
\label{eq1}
\end{equation}
$\hat p$ and $\hat \theta$ are the momentum and angular position operators. The parameters $k$ and $\tau$ represent the laser kick strength and time interval between successive pulses. Considering that experiments occur in a periodic potential, Bloch's theorem may be applied \cite{SW2011}. In our dimensionless variables, the momentum separates into $p=n+\beta$, with an integer part $n$ and a dimensionless quasimomentum $\beta \in [0,1)$. The experimental system's quasi-momentum distribution, primarily determined by the BEC's initial temperature, has typically a width of a few percent in the Brillouin zone.

The evolution over one time period, $\tau$, can be described by the Floquet operator:
\begin{equation}
\hat{\mathcal{U}} = e^{-i\tau \hat{p}^2/2} e^{-ik\sigma_z \cos(\hat{\theta})},
\label{eq2}
\end{equation}
that consists of a free evolution part (left factor) and a kick operator (right factor). For AOKR QWs, the free evolution between subsequent kicks has a duration of a full Talbot time \cite{Dadras2018, Clark2021, dadras2019experimental, Summy2016}, which corresponds to $\tau=\tau_T=4\pi$ in our dimensionless variables \cite{SandroWimberger_2003, SW2011}. Without further nuisance, the free evolution operator is thus equal to unity. 

A kick typically possesses a finite pulse-duration $\tau_p$ itself. Numerically, this is handled by dividing the kick into \( h \) sub-kicks \cite{SandroWimberger_2003}. Each sub-kick is characterized by \(\frac{1}{h}\) of the full kicking strength and is followed by free evolution over \(\frac{1}{h}\) of the total pulse-duration. The parameter \( h \), balancing computational time and approximation accuracy, is chosen based on numerical efficiency; \( h=10 \) has been deemed sufficient for reaching convergence within our parameter regime.
\begin{equation}
\hat{\mathcal{U}}_{\mathrm{kick}} = \prod_{h} e^{-i\frac{\hat{p}^2}{2}\frac{\tau_p}{h}}e^{-i\frac{k}{h} \mathrm{cos}(\hat{\theta})\sigma_z}.  \label{eq3}    
\end{equation}

In this work, the primary concern is spontaneous emission (SE) occurring when the kick potential is on \cite{dadras2019experimental, SandroWimberger_2003, Raizen1999, SW2011}, rather than effects arising from the finite pulse width. Since a finite pulse width effectively introduces reflective walls in momentum space \cite{Raizen1999}, we set a sufficiently small duration for $\tau_p$ to ensure minimal perturbation to momentum-space distributions from this effect.

In resonance conditions, i.e. $\tau=4\pi$, atoms exhibit ballistic movement in momentum space. This means that momentum increases linearly with the number of kicks, revealing the typical signature of a QW \cite{Raizen1999, SW2011, SandroWimberger_2003, Delvecchio2020}. Two internal (spin) states of the atoms are participating in the dynamics. One is moving left, the second one right in momentum space \cite{Summy2016, dadras2019experimental, Dadras2018}, mirrored by the Pauli matrix $\sigma_z$ in Eqs. \eqref{eq2}-\eqref{eq3}. A spin rotation by a so-called coin matrix mixes those two states in each step of the QW \cite{Summy2016, dadras2019experimental, Dadras2018}. In the absence of external perturbations, both internal states would undergo identical evolution due to the kicks. To put it another way, their motion would be symmetric in the walker's space under the influence of the AOKR's evolution operator. This inherent symmetry is broken by imposing a ratchet effect on the initial state, achieved by setting specific initial conditions in the walker's space. To establish this initial ratchet state, we link at least two adjacent momentum classes with a relative phase of $e^{i\pi/2}$:
\begin{equation}
    \ket{\psi_{\rm R}} = \frac{1}{\sqrt{J}}\sum_j e^{i\cdot j\pi/2} \ket{n=j}.
    \label{eq4}
\end{equation}
Such a relative phase shift can be introduced experimentally by the application of conveniently timed Bragg pulses \cite{dadras2019experimental, Dadras2018, Ni2016}. The mean momentum transfer is determined by the sign of the kick potential, which varies between the two internal states. Notably, as the parameter \(J\) in Eq.~\eqref{eq4} increases, the dispersion due to directed kicking decreases. With \(J \geqq 3\), we observe an evolution that most closely mirrors an ideal QW \cite{Ni2016, Bolik2022}
Unless specified differently, our analyses will be based on an initial state with \(J=3\):
\begin{equation}
 \ket{\psi_{\rm R}} = \frac{1}{\sqrt{3}} \left( -i\ket{n=-1}+ \ket{n=0}+i\ket{n=1} \right).
 \label{eq5}
\end{equation}

The final distribution then also depends heavily on the choice of the coin. This coin operation is executed using a Rabi coupling between the atom's two internal states, enabled by resonant microwave pulses, effecting a unitary rotation of these states \cite{dadras2019experimental, PhysRevLett.90.054101, Summy2016}. The QWs presented in Ref.~\cite{Dadras2018,dadras2019experimental,Clark2021} are characterized by a series of unitary operations, as given by:
\begin{equation}
\hat{\mathcal{U}}_\mathrm{step}^t = [\hat{\mathcal{U}}\hat{Q}_1]^t \hat{\mathcal{U}}\hat{Q}_2.
\label{eq6}
\end{equation}
Here, $t\in \bbN$ represents the steps of the walk applied to the initial state. The final momentum distribution's key attributes, like symmetry and form within the walker space, hinge on the specific selection of the initial coin $\hat{Q}_1$ and the coin during the walk protocol $\hat{Q}_2$. We find an asymmetry that emerges due to specific combinations of the coin matrix and a non-resonant evolution of the walk due to the effects of SE. Here, we are especially interested in this asymmetry and in its use. SE is a quantum effect that naturally appears in driven systems. During the application of the laser pulse, the electron in one of the atoms is excited to the energetically higher state and spontaneously collapses into an energetically lower state. This effect is to some extent always present in AOKR experiments \cite{Raizen1999, SW2011, Rev2022} and is typically treated as a nuisance. In this paper, however, we will demonstrate the utility of SE-induced asymmetry for the construction of a QW ratchet \cite{Chakraborty_2017}. 

The shape of the momentum distribution itself will primarily depend on the choice of the coin. Since we intend to analyze the behavior of QWs under the effect of SE, we start in Sec.~\ref{sec-Global_behaviour} with a quick analysis of the coin parameters. Sec.~\ref{sec-SE} is then devoted to the introduction of SE to the system and identifying a candidate coin for engineering a QW ratchet. The key section Sec.~\ref{Sec_Optimization} deals with the optimization of the QW ratchet. Experimental implications are discussed in Sec.~\ref{Sec_experimental}, while Sec.~\ref{sec-conclusion} summarizes our results.

\section{Global behavior of symmetry}
\label{sec-Global_behaviour}

In a recent study, we analyzed discrepancies between experimental data on QWs in momentum space~\cite{Dadras2018,Clark2021,dadras2019experimental} and the corresponding theory. These discrepancies arise from light shift-induced effects on energy levels, which effectively led to the use of a distinct, unintended coin during the walk. This effective coin matrix, when paired with a confined (only two momentum classes with $J=2$) initial state in momentum space, revealed an unusual behavior, leaving a considerable portion of the probability distribution near the origin $n=0$, that vanishes for an initial state with $J \geqq 3$ \cite{Bolik2022}. Such observations suggest that the phase selections within a balanced coin matrix in the quantum-walk protocol can significantly influence key aspects of the final momentum distribution. Here, we are specially interested in the impact of the choice of the coin parameters on the symmetry of the walk.

\subsection{Choice of Coin}
\label{sec-Coin} 

The most general form of the balanced coin matrix for the two-state system with a pseudo spin degree of freedom of the atom is written with three Bloch angles $\chi$, $\gamma$ and $\alpha$ and reads
\begin{eqnarray}
\begin{aligned}
    \hat M( \chi = \frac{\pi}{4}, \gamma, \alpha) &= 
     \begin{pmatrix}
    e^{i \alpha}\cos(\chi= \frac{\pi}{4})            & e^{-i \gamma}\sin(\chi= \frac{\pi}{4})  \\ 
    -e^{i \gamma}\sin(\chi= \frac{\pi}{4}) & e^{-i \alpha}\cos(\chi= \frac{\pi}{4}) \end{pmatrix}\ \\\
    &=\frac{1}{\sqrt{2}}
     \begin{pmatrix}
    e^{i \alpha}          & e^{-i \gamma} \\ 
    -e^{i \gamma} & e^{-i \alpha}\end{pmatrix}.
\end{aligned}
\label{eq7}
\end{eqnarray}
Here $\chi=\frac{\pi}{4}$ determines the coin to be balanced, meaning that both internal states are weighted equally, even though with different phases. Since we are only interested in balanced QWs, we effectively have two free Bloch angles $\gamma$ and $\alpha$ in the coin.

\begin{figure}[!tb]
    \centering
    \includegraphics[width=0.95\linewidth]{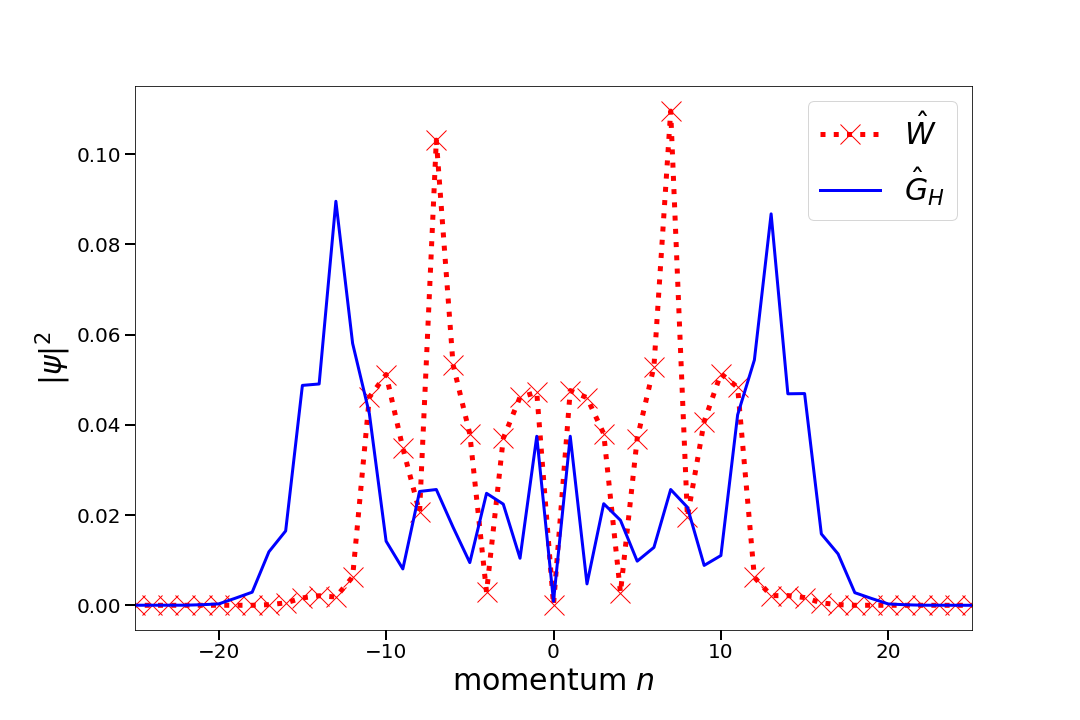}
    \caption{Shown are the momentum distributions after $T=20$ steps in the evolution. The walks are evolved under the $\hat{W}$-coin and the $\hat{G}_H$-coin, respectively.}
    \label{fig1}
\end{figure}

In the following analysis, all walks will be initialized by the so-called $\hat{Y}$-coin \cite{Bolik2022}, choosing in Eq.~\eqref{eq6}
\begin{eqnarray}
    \hat{Q}_1=\hat{Y} = \hat{M}\left(\frac{\pi}{4},\frac{\pi}{2},0\right) = \frac{1}{\sqrt{2}}\begin{pmatrix}
    1 & i  \\ 
    i & 1\end{pmatrix}.
\label{eq8}
\end{eqnarray}
The $\hat{Q}_2$-coin in the walk protocol is then executed by the $\hat{W}$-coin or the $\hat{G}_H$-coin \cite{Bolik2022}:
\begin{eqnarray}
    \hat{W} &=& \hat{M}\left(\frac{\pi}{4},0,0\right) = \frac{1}{\sqrt{2}}\begin{pmatrix}
    1 & 1  \\ 
    -1 & 1\end{pmatrix},
\label{eq9} \\
 \hat{G}_H &=&\frac{1}{\sqrt{2}}\begin{pmatrix}
    1 & 1  \\ 
    1 & -1\end{pmatrix} \widehat{=} \frac{i}{\sqrt{2}}\begin{pmatrix}
    1 & 1  \\ 
    1 & -1\end{pmatrix} \\ &=& \hat{M}\left(\frac{\pi}{4},\frac{3\pi}{2},\frac{\pi}{2}\right) .
\label{eq10}
\end{eqnarray}
Here the equivalence in Eq.~\eqref{eq10} is written, because both coins result in the same walk since a global phase does not induce any change. As discussed in detail in~\cite{Bolik2022}, a walk that evolves under the the $\hat{G}_H$-coin, without any additional distortions, displays markedly different characteristics than a walk evolved under the $\hat{W}$-coin. Notably, there's a pronounced peak near zero momentum, as illustrated in Fig.~\ref{fig1}.

In the following, we will introduce a quantity to characterize the different regimes of symmetry within the parameter regime spanned by the Bloch angles $\alpha$ and $\gamma$.

\subsection{Mapping out parameter space}
\label{sec-Parameterscan}

We are particularly interested in evolving asymmetries in the walk, necessitating the definition of an asymmetry observable $S$. Similarly to \cite{Trautmann_2022}, the value \( S \) is determined by taking the probability to the right of the inversion point at momentum \( n=0 \) and subtracting the probability to the left of it. In the context of Parrondo games two losing strategies combined result in a winning strategy. Using the AOKR-framework for Parrondo-like quantum games \cite{Trautmann_2022}, the sign of the variable $S$ defines the winning or the losing strategy, respectively. In our context, the same measure is subsequently employed as a quantification of symmetry
\begin{eqnarray}
    S = \sum_{n>0}|\psi_n|^2 - \sum_{n<0}|\psi_n|^2.
\label{eq11}
\end{eqnarray}
Here $|\psi_n|^2$ notes the amount of probability associated with momentum class $n$.

In what follows, we will present arguments to shape some intuition and provide qualitative insight into the overall behavior of the observable $S(\gamma, \alpha)$.

\begin{figure}[!tb]
    \centering
    \includegraphics[width=0.95\linewidth]{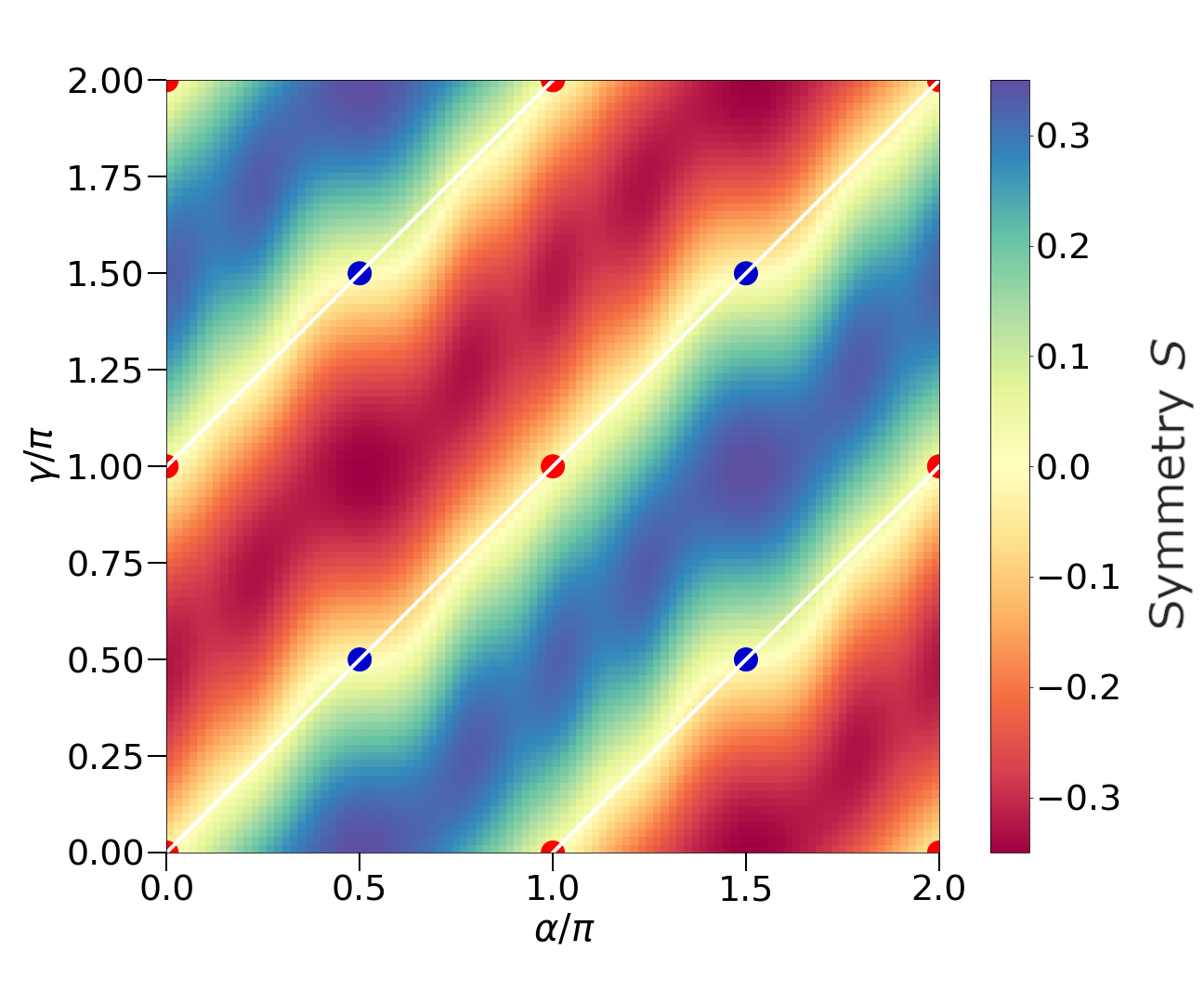}
    \caption{Shown is a colormap of the symmetry indicator $S(\gamma,\alpha)$ in dependency of the free Bloch angles $\alpha$ and $\gamma$, evaluated after $T=15$ steps in the evolution. The blue dots reproduce the walk induced by the $\hat{G}_H$-coin shown in Fig.~\ref{fig1}, while the red dots reproduce the walk induced by the $\hat{W}$-coin. The white lines map out the coin configurations where an additional symmetry is found, investigated in detail in Sec.~\ref{sec-Symmetry}.}
    \label{fig2}
\end{figure}

\subsubsection{Symmetric coin configurations}
\label{sec-Symmetry}

For special relations between $\gamma$ and $\alpha$ additional symmetries can be found in the coin itself. Empirically, this additional symmetry in the coin induces more symmetrical behavior in the walk. Let us discuss two cases. 

For the first case consider $ \gamma=\alpha$, $ \gamma =\alpha \pm 2\pi$:
\begin{eqnarray}
\begin{aligned}
     \hat M( \chi = \frac{\pi}{4}, \gamma, \alpha = \gamma) &=  
    \frac{1}{\sqrt{2}}
     \begin{pmatrix}
    e^{i \alpha}          & e^{-i \alpha} \\ 
    -e^{i \alpha} & e^{-i \alpha}\end{pmatrix}\ \\\
    &\equiv
    \frac{1}{\sqrt{2}}
     \begin{pmatrix}
    \phi          & \rho \\ 
    -\phi & \rho\end{pmatrix} .
\end{aligned}
\label{eq12}
\end{eqnarray}
For the second case consider $\gamma = \alpha  \pm \pi$:
\begin{eqnarray}
\begin{aligned}
     \hat M( \chi = \frac{\pi}{4}, \gamma, \alpha = \gamma\pm\pi) &= 
    \frac{1}{\sqrt{2}}
     \begin{pmatrix}
    e^{i \alpha}          & -e^{-i \alpha} \\ 
    e^{i \alpha} & e^{-i \alpha}\end{pmatrix}\ \\\
    &\equiv
    \frac{1}{\sqrt{2}}
     \begin{pmatrix}
    {\phi}          & -{\rho} \\ 
    {\phi}  & {\rho}\end{pmatrix}.
\end{aligned}
\label{eq13}
\end{eqnarray}
In this setup, we consistently have two pairs of identical phases in the coin. Generally, this condition is not fulfilled. In subsequent computations for each timestep, the internal states are mixed in a comparable manner. It appears plausible that when the states are mixed with similar phases along the protocol, they also evolve in a more balanced way, however systematically in opposite directions. This seems to induce overall a more balanced walk.

In Fig.~\ref{fig2}, the walk evolves for \(T=15\) timesteps, and the symmetry indicator \(S(\gamma,\alpha)\) is plotted as a function of the free Bloch angles. Dominant red or blue stripes lines within the plot mark regions where the QW skews left or right. Intersecting the white lines in-between these areas with \(S=0\) align qualitatively with the anticipated lines in parameter space \(\gamma = \alpha\) and \(\gamma = \alpha \pm \pi\), which possess the coin's additional symmetry from Eq.~\eqref{eq12} and Eq.~\eqref{eq13}, respectively. Not all walks corresponding to $S=0$ in Fig.~\ref{fig2} perfectly match predictions, seemingly distorted by high asymmetry zones, indicating a more intricate dynamic than the rudimentary theory suggests. Still, the theory capably predicts the qualitative behavior of where the balanced walks are found in parameter space due to coin phase symmetry, aiding in understanding for the system. Also the empirical lines with $S=0$ get closer to the predicted lines for a higher number of timesteps, as has been numerically verified.

In Fig.~\ref{fig2}, the blue and red dots are coin types that are identical up to a global phase to \(\hat{W}\) and \(\hat{G}_H\) from Eq.~\eqref{eq9} and Eq.~\eqref{eq10}, respectively. Those points lie on the white lines predicted by our argument from above.

$S$ indicates the direction of evolution for the walk, but not the walk's shape. This means that walks may be balanced, as quantified by $S$, but their shape does not necessarily exhibit an inversion point, underscoring the rarity of true symmetry. Indeed, only two classes of walks exhibit a concrete inversion point.

One class is defined by the \(\hat{G}_H\)-coin walk, presented in Fig.~\ref{fig1}. It is replicated across parameter combinations on the coin-symmetry lines, with half-integer multiples of \(\pi\) for \(\gamma\) and \(\alpha\), indicated by blue full dots in Fig.~\ref{fig2}. Conversely, the \(\hat{W}\)-coin walk is echoed with integer Bloch angle multiples of \(\pi\), shown by red full dots in Fig.~\ref{fig2}. All these walks possess coins that are exclusively real or complex, respectively. When initialized by the \(\hat{Y}\)-coin, they maintain the initial rotation of \(\pi\) in-between the internal states. All other cases introduce both real and imaginary phase components, thus lifting this initial rotation which then appears unfavorable for a mirror-symmetric walk evolution. Thus, other configurations are either balanced without a mirror axis or are distinctly asymmetric.

\subsubsection{Asymmetric coin configurations}
\label{sec-Asymmetry}

In Fig.~\ref{fig2}, the asymmetric regions manifest as broad diagonal stripes, echoing the observations from Sec. \ref{sec-Symmetry}. To investigate the deep asymmetric regime, a broader relationship between $\alpha$ and $\gamma$ can be expressed as $\gamma =\alpha  + \zeta$, where $\zeta$ is arbitrary but constant, giving the coin matrix
\begin{eqnarray}
\begin{aligned}
    \hat M( \chi = \frac{\pi}{4}, \gamma &= \alpha + \zeta, \alpha ) = 
    \frac{1}{\sqrt{2}}
     \begin{pmatrix}
    e^{i (\alpha)}          & e^{-i (\alpha + \zeta)} \\ 
    -e^{i \alpha + \zeta } & e^{-i \alpha}\end{pmatrix}\ \\\
    &\equiv
    \frac{1}{\sqrt{2}}
     \begin{pmatrix}
    {\phi}          & {\rho}e^{-i \zeta} \\ 
    -{\phi}e^{i \zeta}  & {\rho}  \end{pmatrix}.
\label{eq14}
\end{aligned}
\end{eqnarray}

Empirically, we can conclude from Fig.~\ref{fig2} that the walk evolves asymmetrically to the left for $0 < \zeta < \pi$. Thus $e^{i \zeta}$ lives on the right half of the complex unit circle with $\mathrm{RE}(e^{i \zeta}) >0$ and $e^{-i \zeta}$ lives on the left half of the complex unit circle with  $\mathrm{RE}(e^{-i \zeta}) <0$. In comparison, the walk evolves asymmetrically to the right for $-\pi < \zeta < 0$. In this case again $e^{i \zeta}$ lives on the left part of the complex unit circle and $e^{-i \zeta}$ on the respective right half.  
So if the two $\zeta$-related phase angles are effectively interchanged, then the direction in which the walk is skewed is also switched. If this additional $\zeta$-related phase vanishes, then we recover the scenario from Sec.~\ref{sec-Symmetry}.

Looking at the parameter scan from Fig.~\ref{fig2}, we also find spots in the asymmetric parameter-areas, which are exceptionally asymmetric. In the $\mathcal{C}=(\alpha,\gamma)$ coordinate frame, we find that these spots are located at: 
\begin{eqnarray}
     \mathcal{C}_1^+ = (\alpha =  1/2\ \pi , \gamma =  0) \rightarrow \zeta = -\pi/2
\label{eq15}
\end{eqnarray}
\begin{eqnarray}
     \mathcal{C}_2^+ = (\alpha = 3/2 \ \pi , \gamma = \pi) \rightarrow \zeta = -\pi/2
\label{eq16}
\end{eqnarray}
\begin{eqnarray}
     \mathcal{C}_1^- = (\alpha = 1/2 \pi , \gamma =   \pi) \rightarrow \zeta = +\pi/2
\label{eq17}
\end{eqnarray}
\begin{eqnarray}
     \mathcal{C}_2^- =( \alpha =  3/2 \pi , \gamma = 2 \pi) \rightarrow \zeta = +\pi/2
\label{eq18}
\end{eqnarray}

Here the upper index notes that the asymmetry is to the right (+) or to the left (-), respectively. These four points have in common, that they are fully complex on the diagonal and fully real on the off-diagonal elements. With a respective irrelevant phase factor of $i$, this characteristic is similar to the initialization by the coin $\hat{Y}$, which is the same for all walks. This effectively lifts the initial rotation of $\pi$ between the internal states provided by the $\hat{Y}$-coin and causes the walk to be highly asymmetric.

While the results oft the section may be interesting for designing Parrondo-like games \cite{Trautmann_2022}, we are in the following interested in how SE affects the symmetry of the QW.

\section{spontaneous emission}
\label{sec-SE}

So far, our focus has been on asymmetries in the coherent QW stemming from the selection of the coin matrix. However, noise-induced asymmetries garner particular attention, given their potential utility in crafting a QW ratchet \cite{Chakraborty_2017}. The QW ratchet described in this study is particularly noteworthy as it is induced by SE. Typically, SE is associated with merely causing decoherence \cite{Clark2021}, but in this case it contributes to the creation of a QW ratchet. This phenomenon represents a shift in perspective, where what is usually considered a nuisance becomes advantageous.

\subsection{Spontaneous emission in the AOKR}

In the context of AOKR-dynamics a SE event can be interpreted as an event in which the atom goes into an excited state during the application of the laser-pulse and then spontaneously falls either into the energetically lower $|1\rangle$ or $|2\rangle$ state. All excitations are assumed to be far away from an internal electronic resonance, in contrast to \cite{Andersen2020, Andersen2022}.
During the spontaneous de-excitation a photon is emitted, inducing also a shift of the atomic quasimomentum \cite{SandroWimberger_2003, SW2011}.
For the AOKR we arrive at the following effective Hamiltonian, for which a detailed derivation can be found in chapter 5 of \cite{groiseau2017discrete} and \cite{Caspar2019} for non-resonant driving
\begin{eqnarray}
\begin{aligned}
    \hat{\mathcal{H}} &= \frac{1}{1+\frac{\mu^2}{4\delta^2}} \frac{\Omega_1^2}{8\delta} \big(\mathrm{cos}(2k_L\hat{x})+1\big) |1\rangle\langle1| \ \\\
    &-\frac{1}{1+\frac{\mu^2}{4\Delta^2}} \frac{\Omega_2^2}{8\Delta} \big(\mathrm{cos}(2k_L\hat{x})+1\big) |2\rangle\langle2|.
\end{aligned}
\label{eq19}
\end{eqnarray}

\begin{figure}[!tb]
    \centering
    \includegraphics[width=0.95\linewidth]{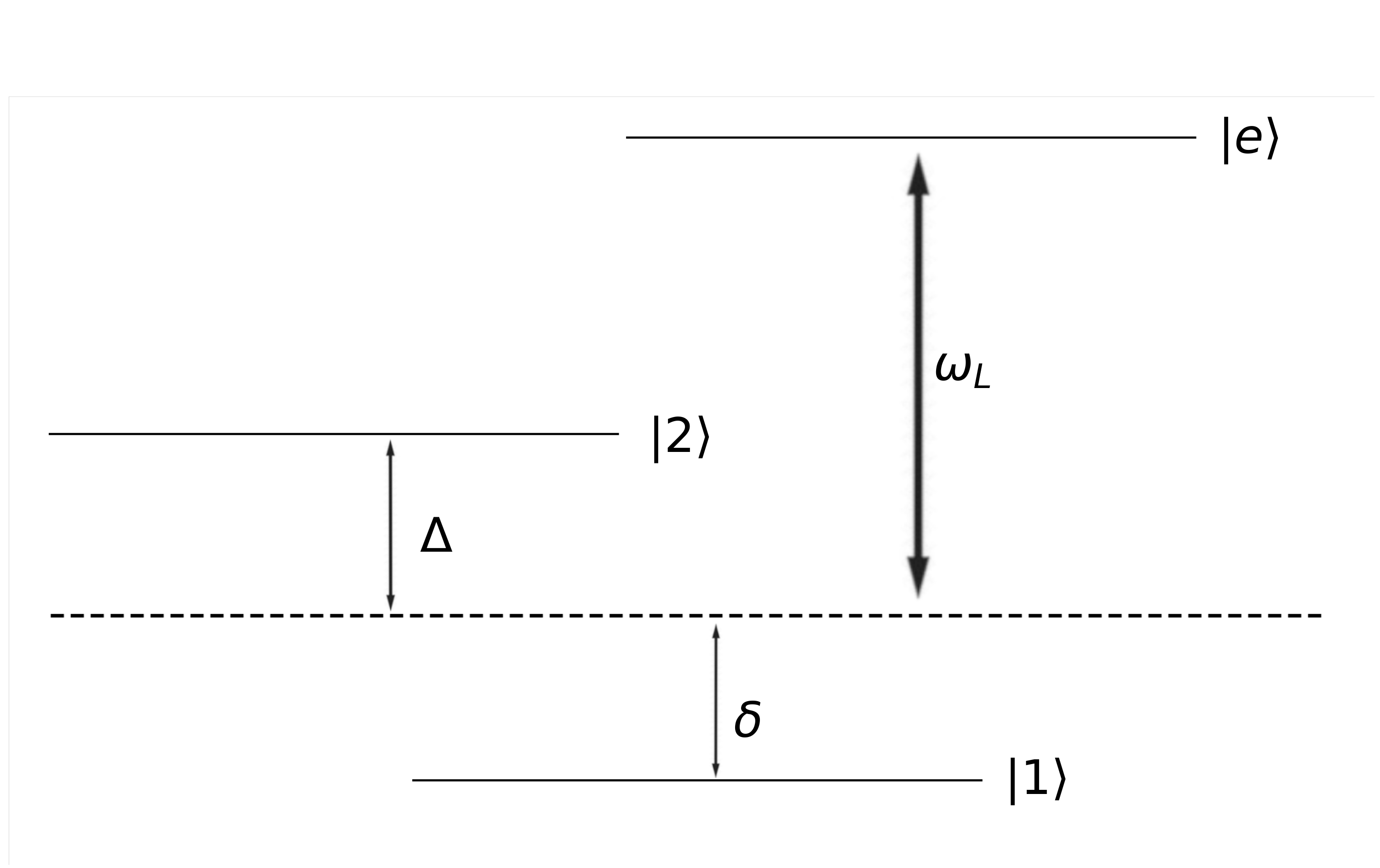}
    \caption{Level scheme of the AOKR quantum walk with the two internal levels $\ket{1}$ and $\ket{2}$. The excited state $|e\rangle$ is used to impart the kicks to the atoms by a off-resonant standing wave of laser light. $\omega_L$ denotes the frequency of the kicking laser. $\delta$ and $\Delta$ denote the detunings of the two possible transitions with respect to the laser frequency.}
    \label{fig3}
\end{figure}

Let \(\mu\) denote the SE event rate, defined as:
\begin{equation}
\mu = \mu_1 + \mu_2 = \frac{k_1}{\tau_p\tau_{\rm SE}|\delta|} + \frac{k_2}{\tau_p\tau_{\rm SE}|\Delta|}    
\end{equation}
\label{eq20}
where \(\delta\) and \(\Delta\) are the respective detunings from the resonance, shown in Fig.~\ref{fig3}. Given that the detuning is small compared to the laser frequency, the approximation \(k_1 = k_2 = k\) is valid \cite{Wimberger2016, dadras2019experimental}. Consequently we have $\delta = -\Delta$ and thus we find $\mu_1 = \mu_2$ for the SE rates in Eq.~\eqref{eq20}.
Also, \(\tau_p\) is the pulse duration, and \(\tau_{\rm SE}\) represents the lifetime of the transition.

In a numerical context, a SE event translates into a projection onto one of the internal states:
\begin{align}
\hat{P}_{|1\rangle} &= \frac{1}{\mathcal{N}} \left(A |\psi_{|1\rangle}\rangle \otimes |1\rangle \langle1| + B |\psi_{|2\rangle}\rangle \otimes |1\rangle \langle 2|\right), \label{eq21} \\
\hat{P}_{|2\rangle} &= \frac{1}{\mathcal{N}} \left(C |\psi_{|1\rangle}\rangle \otimes |2\rangle \langle1| + D |\psi_{|2\rangle}\rangle \otimes |2\rangle \langle 2|\right), \label{eq22}
\end{align}
accompanied by a shift in quasimomentum, \(\beta_{\rm SE} \). Here, \(\mathcal{N}\) serves as a normalization factor and \(A, B, C, D\) are detuning-dependent weights. For our analysis, \(A,B,C,D\) can be approximated as \(1/2\), see \cite{Caspar2019}.

With the inclusion of SE, the kick's evaluation becomes more complex \cite{SW2011, SandroWimberger_2003}. For each of the \(h\) sub-kicks from Eq.~\eqref{eq3}, we determine, based on an event probability \(p_{\mathrm{event}} = \frac{\mathrm{p_{SE}}}{h}\), whether an SE event transpires. If it does, with a 50-50 chance we decide the projection onto either \(\hat{P}_{|1\rangle}\) or \(\hat{P}_{|2\rangle}\) internal states, and the quasimomentum shift is randomly selected from \(\beta_{\rm SE} \in \mathrm{uniform}(-0.5, 0.5)\). In the event's presence, the partial kick is expressed as:
\begin{eqnarray}
\hat{\mathcal{U}}_{\mathrm{SE}}=\hat{P}_{|i\rangle}e^{-i\frac{(\hat{p}+\beta_{\rm SE})^2}{2}\frac{\tau_p}{h}}e^{-i\frac{k}{h} \mathrm{cos}(\hat{\theta})\sigma_z},
\label{eq23}
\end{eqnarray}
with \(|i\rangle\) being \(|1\rangle\) or \(|2\rangle\), with a probability of $1/2$. The Pauli matrix $\sigma_z$ incodes that one internal state is moving left, the other one right due to the kick. Our approach concurrently accounts for the kick's finite duration and the potential of multiple events during a single pulse duration. 
For $^{87}\mathrm{Rb}$ quantum walks \cite{dadras2019experimental,  Clark2021, Dadras2018}, the Talbot time is $103\mu$s, equivalent to \(4\pi\) in our units. In AOKR experiments, the typical pulse width is of the order of $100$ns, but much shorter pulses should be possible, see e.g. \cite{Ryu2006, Yingmei}. We use here throughout $\tau_p = 0.005$, corresponding to a pulse duration of a few tens of ns. This minimizes the impact of possible perturbations due to the finite pulse width \cite{PhysRevLett.75.4598, Klapp1999, Fishman1986, SandroWimberger_2003} for longer walks of more than a few tens of steps.

The total kick is then again written as: 
\begin{equation}
\hat{\mathcal{U}}_{\mathrm{SE-kick}} = \prod_{h} \hat{\mathcal{U}}_{\mathrm{SE}}.  \label{eq24}    
\end{equation}
The shift in quasimomentum enters in the free evolution in-between subsequent kicks with a duration of a full Talbot period in the same way. \begin{equation}
\hat{\mathcal{U}}_{\mathrm{SE-free}} = e^{-i\frac{(\hat{p}+\beta_{\rm SE})^2}{2}\tau_T} 
\label{eq25}    
\end{equation}
Since the shift in quasimomentum comes from a uniform distribution and the Talbot period is with a duration of $\tau_T=4\pi$ relatively large compared to the duration of the pulse itself, the shift in quasimomentum is of a much greater significance here, then during the kick pulse in Eq. \eqref{eq23} \cite{SW2011, SandroWimberger_2003}. Indeed, the free evolution between subsequent pulses with the duration ${\tau_T = 4\pi}$ but off-resonant shift in quasimomentum due to SE, can lead to strongly non-ballistic (nonresonant) diffusive evolution \cite{Clark2021, groiseau2017discrete} that effectively destroys the QW.

has been identified as a strong nuisance due to their long duration. Mainly, this effect limits the regime of event rates in which SE-induced effects will be observable, which will be discussed in more detail in the following analysis.

\subsection{Spontaneous emissions during quantum walk}
\label{sec-SEintroduction}

\begin{figure}[tb]
    \centering
    \includegraphics[width=0.95\linewidth]{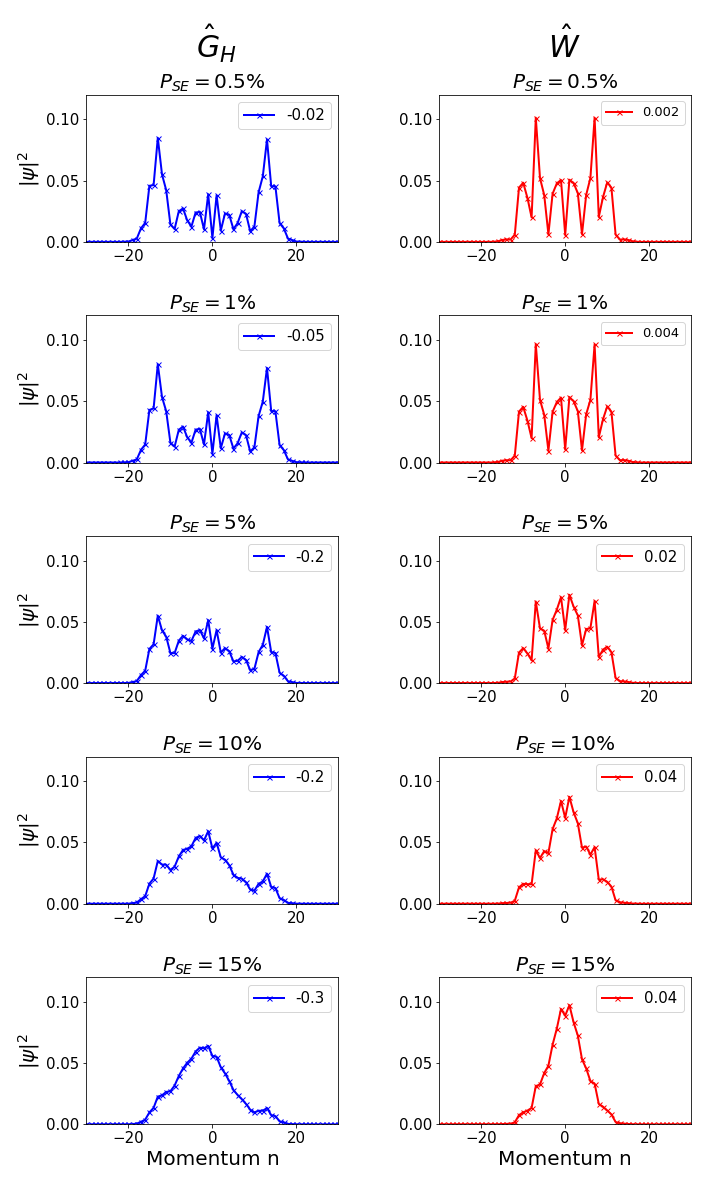}
    \caption{Two different walk protocols evaluated after $T=15$ steps in the evolution with increasing SE. Both protocols are initialized by the $\hat{Y}$-coin. The left row is then evolved under the $\hat{G}_H$-coin, while the right row is evolved under the $\hat{W}$-coin. From top to bottom SE rates increase. The number in the upper right corner in each plot indicates the symmetry of the evolution $S$. The $\hat{G}_H$-walk develops a strong asymmetry while the $\hat{W}$-walk becomes only slightly asymmetric. It might be suspected that the weak asymmetry in the $\hat{W}$-protocol is due to numerical fluctuations. However, it appears to be relatively robust. The distributions present an average over $10000$ realizations, with the given probabilities of $p_{\rm SE}$ per kick.}
    \label{fig4}
\end{figure}

From Fig.~\ref{fig4}, we observe that the walk under the \(\hat{W}\) coin, with increasing SE rates, becomes more classical and exhibits mild asymmetry. This can be read off $S$ which is shown in the top right corner of each subplot. Conversely, the walk under the \(\hat{G}_H\) coin displays a quite pronounced asymmetry, always a magnitude larger when compared to the $\hat{W}$ walk. 
For high event probabilities, the ballistic peaks for positive momentum classes are noticeably damped by SE. However, the peaks for negative momentum classes are less affected. As SE rates increase, the walk becomes more classical and the effect blurs out. 

\begin{figure}[tb]
    \centering
    \includegraphics[width=0.75\linewidth]{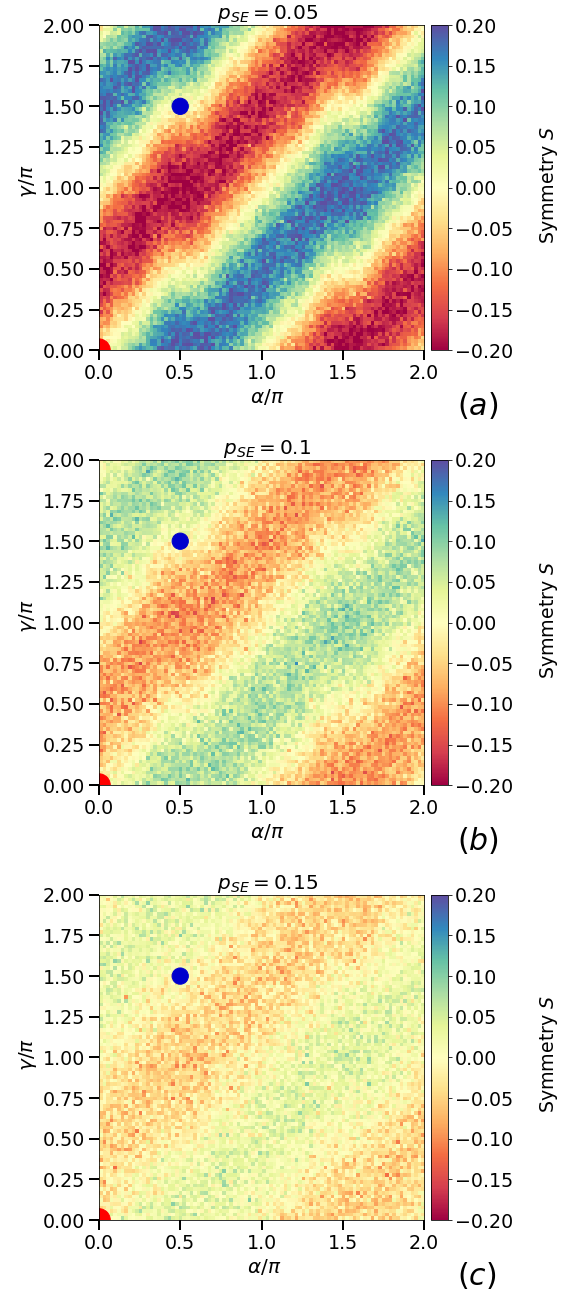}
    \caption{Scan of the symmetry $S(\gamma,\alpha)$ of the QW, evaluated after $T=15$ steps in the walk. The walk is initialized by the $\hat{Y}$-coin and then evolved under ${\hat{M}(\chi= \frac{\pi}{4},\gamma,\alpha)}$. From top to bottom SE rates are increased. The red dot at $(\gamma=0,\ \alpha=0)$ marks the more symmetric walk evolved under the $\hat{W}$-coin, while the blue dot at $(\gamma=1.5\pi,\ \alpha=0.5\pi)$ marks the asymmetric walk evolved under the $\hat{G}_H$-coin (compare to Fig.~\ref{fig2}). It appears that the $\hat{G}_H$-walk lives in a more unstable area in parameter space with respect to the symmetry of the evolution under the influence of SE, causing an initially symmetric walk to evolve asymmetrically under the introduction of SE. Each pixel is calculated as an average over 1000 event trajectories.}
    \label{fig5}
\end{figure}

With rising SE rates, the diagonal structures within the free Bloch angles parameter domain blur, as illustrated in Fig.~\ref{fig5}. 
In this parameter landscape, the \(\hat{W}\)-walk (red dot) maintains a position where even at high SE-rates, only a small deviation from the initial parameter scan from Fig.~\ref{fig2}, resulting in minor asymmetries, also consistent with Fig.~\ref{fig4}. In contrast, the \(\hat{G}_H\)-walk (blue dot) evolves strong asymmetries as it lies in a region dominated by the pronounced asymmetry. This is due to increasing SE rates, the asymmetric regions, governed by \(\mathcal{C}_i^\pm\), shift towards larger values of \(\gamma\). This shift now situates the \(\hat{G}_H\) coin within the asymmetric region. This finding opens a route to design an open QW ratchet.

Within the parameter space of $\gamma$ and $\alpha$, SE introduces two key alterations: a transition from a strong stripe to a more faint pattern, followed by a drift of the asymmetric regions to larger values of \(\gamma\). 
The walk exhibiting an asymmetry then depends on the coin's initial position within this parameter space. The effects are counteracting each other. The shift in $\gamma$ situates the $\hat{G}_H$-walk in a region of higher asymmetry vs. the fainting amplitude of asymmetry as a consequence of the off-resonant walk. 
Thus, depending on the choice of coin an asymmetric evolution of the walk can be constructed with a corresponding SE rate.
In other words, for given windows of $p_{\rm SE}$, the initially symmetric QW can be engineered to be maximally asymmetric.

Numerically, SE are implemented by two distinct effects: (1) the projection onto an individual internal state with a 50:50 probability, and (2) the shift in quasimomentum. We studied each of these effects separately for a better understanding. Focusing solely on the projection with resonant quasimomentum, we observe a shift in the stripe pattern in seen in Fig.~\ref{fig2} towards higher values of $\gamma$. This shift becomes pronounced at SE rates of approximately $20\%$, though it is also present at lower rates in a diminished form. The $\hat Y$-Coin is located at $(0,0)$, which, within this scenario, represents a particularity since the white line with $S=0$ always extrapolates to the origin $(0,0)$. Hence this point is not effected by the shift. The other coin, $\hat{G}_H$, is positioned at $(3/2\pi, 1/2\pi)$ and finds itself in a region with $S \neq 0$ due to the shifting of the stripes. Adding (2) the shift in quasimomentum present in the numerical model of SE, the overall image becomes noticeably noisier. This is attributed to the new, shifted quasimomentum being drawn from a broad random distribution, hiding the effect of the projection. Unfortunately, with both SE-induced effects present, the clarity of the stripe shift in Fig. \ref{fig5} is compromised. The combination of these effects leads to a washing out of the data for $p_{\rm SE} \geq 20\%$ SE rates, which motivates the further study of $p_{\rm SE} \leq 15\%$. Overall we can say that the SE-induced shift of quasimomentum, making it in particular non-resonant, together with the right choice of the coin induces the asymmetry in the QW effected by SE.

\section{Optimizing the QW ratchet}
\label{Sec_Optimization}

The SE-induced asymmetry observed for the $\hat{G}_H$-walk is intriguing since it may be used for the construction of a ratchet. This section analyzes in detail such an engineering of an AOKR-ratchet in the quantum- and as well in the classical regime. 

Previous AOKR ratchets were so-called Hamiltonian ratchets \cite{Ni2017, Sadgrove07, PhysRevLett.101.180502}, where the symmetry is broken by the initial conditions, while the evolution remains coherent. 
This raises curiosity about the optimal parameter choice to maximize the asymmetry of the $\hat{G}_H$-walk and thereby enhance the efficacy of an SE-induced ratchet.
For our analysis we will only consider the walk of the $\hat{G}_H$-coin, since unperturbed it has an inversion point. Therefore, it is a symmetric walk without SE that turns asymmetric only when perturbed by SE. Thus our AOKR system is a suitable candidate for a purely SE-induced ratchet.

\subsection{Initial state dependence}

Any AOKR QW \cite{dadras2019experimental, Dadras2018, Ni2016, Delvecchio2020, Summy2016, Clark2021, Bolik2022} is built on a coherent Hamiltonian ratchet and hence on the localization of the initial conditions within the kicking potential \cite{Ni2017, SW2011, Ni2016, Sadgrove07}.
The width of the initial state in momentum space and thus the amount of momentum classes included in the initial state $J$, as in Eq.~\eqref{eq4}, holds significant relevance for the shape of the distribution \cite{Ni2016}. As discussed in \cite{Bolik2022}, a broader initial state considerably diminishes the portion of the distribution proximal to $n=0$ for the $\hat{G}_H$-walk. Consequently, a broader initial state implies diminished central probability, which is instead prevalent in the tails, amplifying the asymmetry. 

\begin{figure}[tbp]
    \centering
    \includegraphics[width=\linewidth]{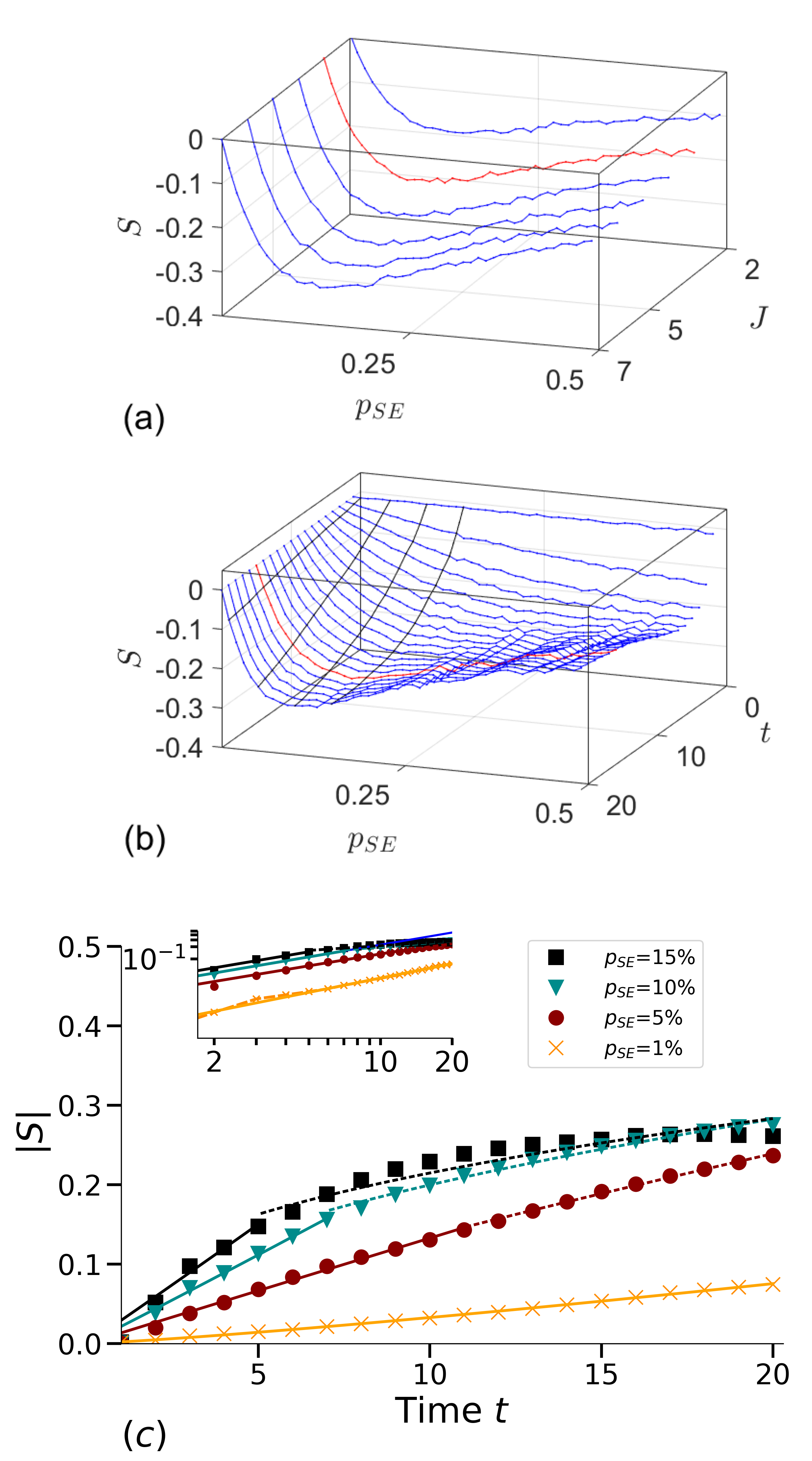}
    \caption{Dependency of the symmetry observable $S(p_{\rm SE})$ on the probability per kick $p_{\rm SE}$ for a SE event. Panel~(a): $S(p_{\rm SE})$ for the walk conducted by the $\hat{G}_H$-coin and observed after $T=15$ steps with an initial state of increasing width, as described by Eq.~\eqref{eq4} and denoted by $J$. Panel~(b): $S(p_{\rm SE})$ with increasing numbers of timesteps $t$. Here the initial state is conducted by $J=3$ momentum classes. The red curves show the system for three momentum classes in the initial state and observed at $t=15$ steps in the evolution, as also discussed in Fig.~\ref{fig1}. 
    The black curves represent $S(p_{\rm SE},t)$ for fixed $p_{\rm SE} = \ 1\%,\ 5\%,\ 10\%,\ 15\%$. These curves are shown in panel~(c) with their corresponding power-law fits with exponents:
    $a=1$ for $p_{\rm SE} = 1\%$ for the entire time range, crossover from $a=1$ towards $a=0.5$ for $p_{\rm SE} = 5\% \dots 15\%$. With increasing SE rates, the quantum regime with $a=1$ (straight line fits) becomes shorter and shorter with respect to the crossover part (dotted lines).
    The curves and their respective fits are also shown in an inset on a double logarithmic scale.
    Data is obtained from averages over 10000 trajectories modeling SE \cite{groiseau2017discrete, Caspar2019}. }
    \label{fig6}
\end{figure}

The walk we are analyzing exhibits perfect symmetry for $p_{\rm SE} = 0$. At elevated SE probabilities, the walk reverts to a classical system due to strong phase decoherence. This suggests the existence of an optimal SE probability that maximizes the asymmetric walk evolution. The argument holds within the region in which the walk is not yet dominated by classical dynamics, which would be the case around $p_{\rm SE}=10\%$ and below.
Our hypothesis is supported by Fig. \ref{fig6}~(a) which shows the dependency of the symmetry observable $S$ on $p_{\rm SE}$ and on $J$.
We find that the curve of the symmetry indicator $S(p_{\rm SE}, J)$ improves with increasing momentum classes $J$ in the initial state and maximizes for $T=15$ at $p_{\rm SE} \approx 0.1$. However, from Fig.~\ref{fig4} we know that the regime up to which the resonant tails of the distribution are still visible lies roughly around $p_{\rm SE} \approx 0.05$, which would be the regime in which a ratchet that exhibits quantum resonance \cite{Raizen1999, SW2011, SandroWimberger_2003} might be well observable.

The biggest improvement on the asymmetry comes, when increasing from an initial state constructed with $J=2$ to $J=3$. This observation seems to align with the results from \cite{Bolik2022}, since for $J=3$ the part of the momentum distribution located around $n=0$ significantly decreases and is thus found in the tails, contributing more to the asymmetry. A similar argument can be made when increasing from $J=3$ to $J=4$. For broader initial states the improvement becomes less significant. The better the localization in angular $\theta$ space, the better the walks evolve with minimal dispersion. 
In our simulations, we have typically chosen $J=3$ since it seems as a good tradeoff between experimental feasibility and the amplitude of asymmetry.
Also note that the asymmetries do not fully revert to zero, even far in the classical regime.

\subsection{Time dependence of the asymmetry}

The temporal evolution of asymmetry offers insights into reasonable timescales for observing this effect experimentally. In Fig.~\ref{fig6}~(b), $S(p_{\rm SE},t)$ reveals how this unfolds. Notably, the red curve represents the state at $t=15$, as previously elaborated. 
The black curves delineate the asymmetry's temporal development, $S(p_{\rm SE},t)$, at specific $p_{\rm SE}$ values. The measurement of $S$ is connected to the speed of the expansion of the walks, which clearly depends on $p_{\rm SE}$. The signature of QWs is their ballistic expansion, which would correspond to an exponent of $a=1$ for a power-law fit of $S(t) \propto t^a$.  Conversely, a classical walk showcases a diffusive expansion, signified by $a=1/2$. As Fig.~\ref{fig6}~(c) suggests, at a marginal $p_{\rm SE}=1\%$, the expansion remains entirely within the quantum regime, exhibiting $a=1$. As $p_{\rm SE}$ increases, we discern a progression from a purely quantum expansion ($a=1$) to one that is more classical with a diminished exponent $a$. Notably, at elevated $p_{\rm SE}$ values, the trend predominantly aligns with classical behavior, approaching an exponent of $1/2$. Also note that for progressing timesteps $t$ the maximal observable asymmetry is visible at smaller values of $p_{\rm SE}$. This effect is mainly caused by the off-resonant quasimomenta in the free evolution between subsequent pulses.

For event probabilities far in the classical regime the walk exhibits an asymmetry that initially grows fast and then reverts due to strong decoherence as time proceeds. This is especially present in the black curve in Fig.~\ref{fig6}, corresponding to an event probability of $p_{\rm SE}=15\%$. In this case, not discussed in further detail, the off-resonant quasimomenta dominate the dynamics and asymmetries effectively decline as time proceeds.

Essentially, all this demonstrates a transition from quantum to classical behavior. Augmented decoherence — either due to pronounced $p_{SE}$ values or a combination of moderate $p_{\rm SE}$ with extended durations — renders the walk diffusive. Three distinct regimes emerge, identified by the best parameters from a powerlaw fitting:\\
(i) The pure quantum regime, evidenced at $p_{\rm SE}=1\%$.\\
(ii) An intermediate regime exhibiting a quantum-to-classical transition as $t$ increases, seen at ${p_{\rm SE}=5\%,10\%}$.\\
(iii) A predominantly classical regime, as evident for $p_{\rm SE}=15\%$ and above.
This observation is in agreement with Fig.~\ref{fig4}, where we identified the diminishing resonant tails in the momentum distribution for $p_{\rm SE}=15\%$.

\section{Experimental constraints}
\label{Sec_experimental}

Typical experiments start with a Bose-Einstein condensate centered around $n=0$, but with a distribution in quasimomentum of finite width \cite{Dadras2018, dadras2019experimental, Ryu2006}. 
In-between the pulses and during the pulse itself, this nonzero quasimomentum causes non-resonant phase effects \cite{SW2011, SandroWimberger_2003}. Thus, similar to the influence of SE, quantum resonances are perturbed by non-resonant $\beta$´s.
While up to now, the initial quasimomentum was chosen to be vanishing $\beta_{\mathrm{init}}=0$, now we consider for our simulations a Gaussian distribution of width $\beta_{\rm FWHM}$ modeling experimental reality \cite{Dadras2018, dadras2019experimental, SW2011, Ryu2006}.

\begin{figure}[tbp]
    \centering\includegraphics[width=0.75\linewidth]{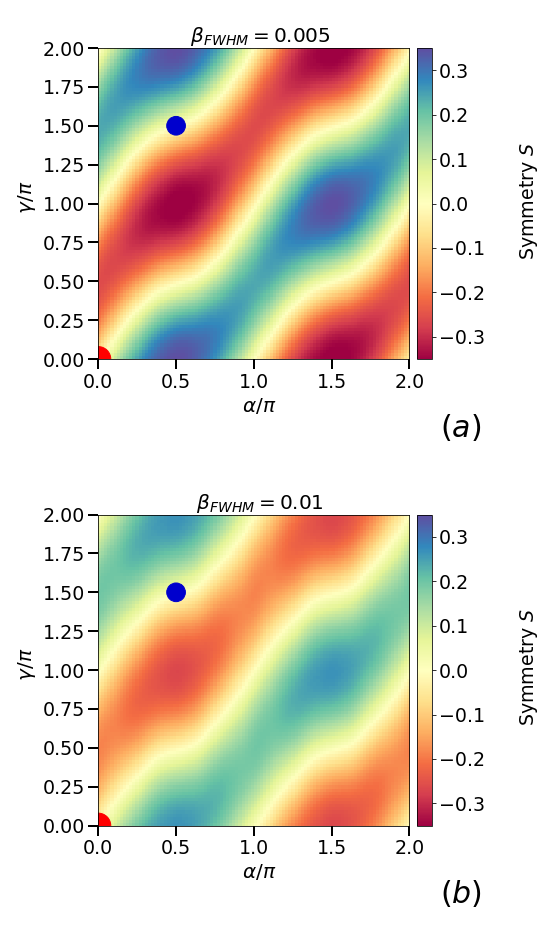}
    \caption{$S$ as a function of the free Bloch angles $\alpha$ and $\gamma$ under increasing width of the initial quasimomentum distribution. $S$ is evaluated after $T=15$ steps in the evolution of the walks. The quasimomentum is drawn from a Gaussian distribution with width $\beta_\mathrm{FWHM}$. Each point in the parameter scan is averaged over 1000 trajectories. The red dot at $(\gamma=0,\ \alpha=0)$ and the blue dot at $(\gamma=1.5\pi,\ \alpha=0.5\pi)$ mark the parameters corresponding to the $\hat{W}$-walk and the $\hat{G}_H$-walk, respectively. 
    }
    \label{fig7}
\end{figure}

Similarly to the introduction of SE from Sec.~\ref{sec-SEintroduction}, a finite quasimomentum distribution causes the asymmetric areas to bulk around the points of high asymmetry $\mathcal{C}_i^\pm$, as demonstrated in Fig.~\ref{fig7}~(a) and (b). A similar deviation from the original stripe pattern has been noted In Sec.~\ref{sec-SEintroduction}. For broader quasimomentum distributions, see Fig.~\ref{fig7}~(b), the asymmetries become weaker and the walks become more diffusive. However, in contrast to the perturbations caused by SE, there is no additional shift of the asymmetric regions to higher $\gamma$. Thus
the walks calculated by the $\hat{G}_H$-coin and the $\hat{W}$-coin do not evolve new strong asymmetries. Overall they just become more "classical"-like in the sense that they show a diffusive instead of a ballistic evolution. Thus the shift noticed in the asymmetric regions of Fig.~\ref{fig5} seems to be an effect unknown to other typical forms of perturbations in the AOKR-framework.

However, the qualitative behavior of $S(\gamma,\alpha)$ under the influence of a non-resonant quasimomentum distribution is similar to the non-resonant evolution found under the influence of SE.

\begin{figure}[tbp]
    \centering\includegraphics[width=0.9\linewidth]{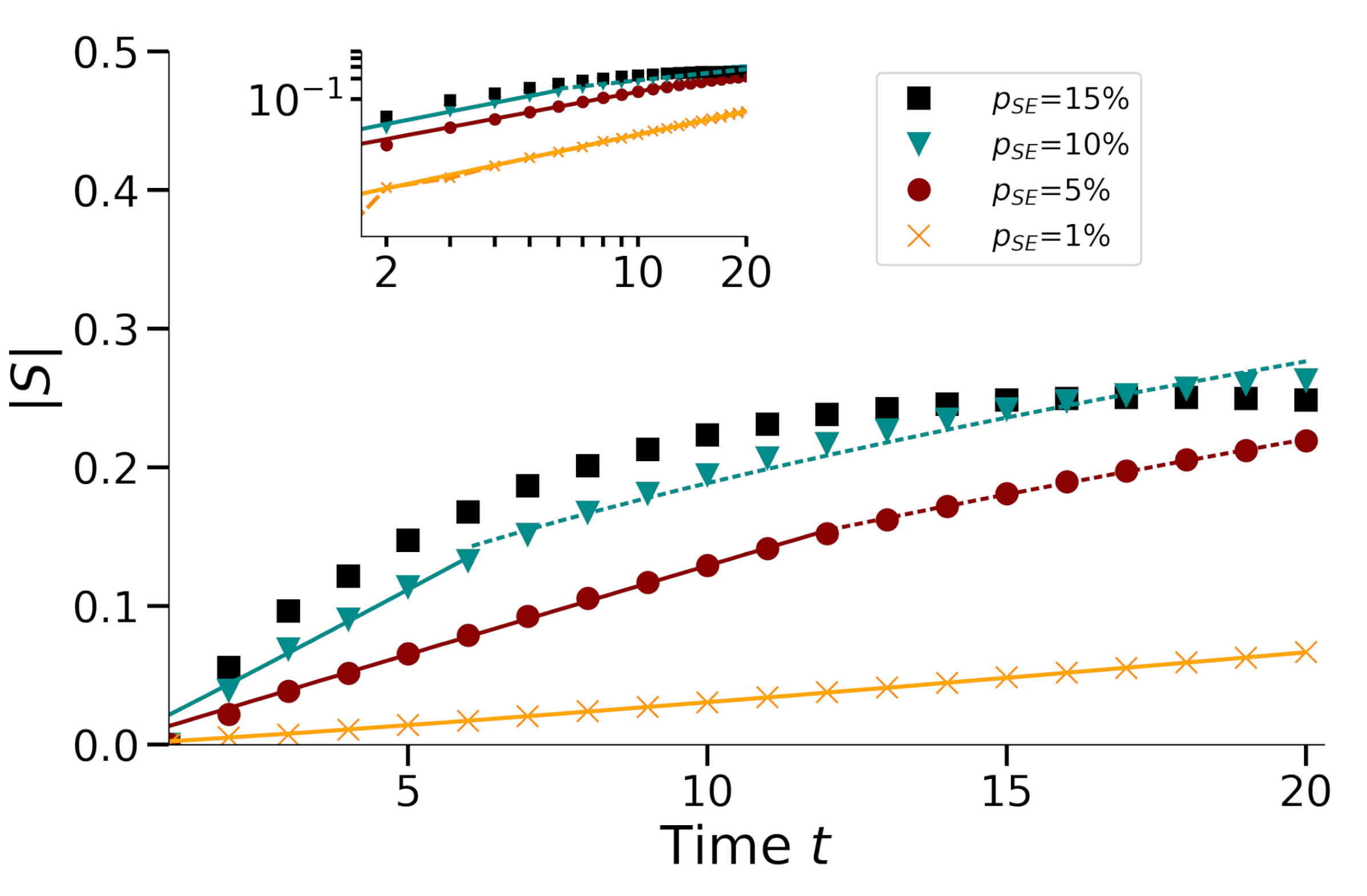}
    \caption{Shown is the time dependence of the symmetry observable $S$.  Considered are the combined effects of finite quasimomentum with $\beta_{\rm FWHM}=0.01$ and SE. The individual curves showcase several different values of ${p_{\rm SE}=1\%,\ 5\%,\ 10\%,\ 15\%}$. The straight lines mark out the regime that results with an exponent of $\alpha=1$ in a power-law fit, while the dotted areas mark the regime with $\alpha< 1$. The curves are evaluated as an average of 200 initial quasimomenta, each realized with 1000 SE-trajectories.
    }
    \label{fig8}
\end{figure}

It has been numerically verified that the walk does indeed demonstrate no significant asymmetries over the range of 20 timesteps under the influence of solely finite quasimomentum and without SE. In Fig.~\ref{fig8}, we see the time dependence of $S$ for several different values of $p_{\rm SE}$, similar to Fig.~\ref{fig6}~(c), however with additional quasimomentum of $\beta_{\rm FWHM}=0.01$. 
The figure indicates that the effect survives also with an additional quasimomentum, which would be present in an experimental setup. Again we observe qualitatively similar to Fig.~\ref{fig6}~(c) the quantum to classical transition with three different regimes, corresponding (i) to a pure quantum mechanical regime, (ii) a regime that starts ballistic and then evolves a diffusive motion as decoherence increases over time, and (iii) the predominantly classical regime. 
The finite initial quasimomentum introduces additional dispersion, which sets the expectation that the transition should settle in earlier. However, this is only partially true. 
For higher event probabilities $p_{\rm SE} = 10\%$,  we find that the transition settles in for shorter times, which would also be expected due to higher dispersion in the system. This additional dispersion slightly shifts the borders of the quantum mechanical regime (ii) with respect to the case of Fig.~\ref{fig6}.
It is worth mentioning that for a hypothetical future experiment, too large pulse durations for $\tau_p$ would effectively induce barriers in momentum space \cite{Raizen1999,PhysRevLett.75.4598, OSKAY2003409}. However, a QW conducted with a $^{87}\mathrm{Rb}$ BEC has already been found to be observable on the discussed timescales of up to 20 timesteps \cite{Dadras2018,Clark2021,dadras2019experimental, Bolik2022}.

\section{Conclusion}
\label{sec-conclusion}

Our study has elucidated insights into the intricate interplay between symmetry and the pivotal role of the choice of the coin and the impact it has on the system's overall dynamics. The introduction of SE has proven instrumental as a mechanism that introduces asymmetry into an initially symmetric QW for the construction of a QW ratchet. 

Typically, SE-induced effects are detrimental. However, SE manifests indeed beneficially under certain conditions \cite{Oxford2001, Andersen2020, SandroWimberger_2003}. Our observations on the temporal behavior of the symmetry observable $S$ chart a distinct transition from quantum to classical behavior depending on the strength of the introduced SE. Evidence for this is found by an power-law fit to $S(t)$ thereby observing a transition from a ballistic to a sub-ballistic expansion, corresponding to the quantum to classical transition in the walk. The origin of the observed asymmetry is an interference effect arising from specific choices of the coin matrix and the non-resonant free evolution due to the shifts in quasimomentum induced by SE. 

Our findings have proven to be robust under the influence of finite initial quasimomentum, which would be present in an experimental setup due to finite temperature. The resilience of the observed effects under this perturbation underscores the feasibility for the experimental realization of an SE-induced QW ratchet.

Overall, this investigation offers a deeper, more nuanced understanding of the dynamics of AOKR QWs and the influence of SE on its dynamics. By delineating the unexpected utility of SE for an AOKR ratchet and highlighting the robustness of our system against quasimomentum, our work provides insight for potential future experiments along the line of \cite{dadras2019experimental,  Clark2021, Dadras2018, Trautmann_2022}. 

\vspace{5mm}

\begin{acknowledgments}
We are grateful to Caspar Groiseau for help on the implementation of SE simulations.
S.W. acknowledges funding by Q-DYNAMO (EU HORIZON-MSCA-2022-SE-01) with project No. 101131418 and by the National Recovery and Resilience Plan, Mission 4 Component 2 Investment 1.3 -- Call for tender No. 341 of 15/03/2022 of Italian MUR funded by NextGenerationEU, with project No. PE0000023, Concession Decree No. 1564 of 11/10/2022 adopted by MUR, CUP D93C22000940001, Project title ``National Quantum Science and Technology Institute".
\end{acknowledgments}


\begin{thebibliography}{51}%
\makeatletter
\providecommand \@ifxundefined [1]{%
 \@ifx{#1\undefined}
}%
\providecommand \@ifnum [1]{%
 \ifnum #1\expandafter \@firstoftwo
 \else \expandafter \@secondoftwo
 \fi
}%
\providecommand \@ifx [1]{%
 \ifx #1\expandafter \@firstoftwo
 \else \expandafter \@secondoftwo
 \fi
}%
\providecommand \natexlab [1]{#1}%
\providecommand \enquote  [1]{``#1''}%
\providecommand \bibnamefont  [1]{#1}%
\providecommand \bibfnamefont [1]{#1}%
\providecommand \citenamefont [1]{#1}%
\providecommand \href@noop [0]{\@secondoftwo}%
\providecommand \href [0]{\begingroup \@sanitize@url \@href}%
\providecommand \@href[1]{\@@startlink{#1}\@@href}%
\providecommand \@@href[1]{\endgroup#1\@@endlink}%
\providecommand \@sanitize@url [0]{\catcode `\\12\catcode `\$12\catcode
  `\&12\catcode `\#12\catcode `\^12\catcode `\_12\catcode `\%12\relax}%
\providecommand \@@startlink[1]{}%
\providecommand \@@endlink[0]{}%
\providecommand \url  [0]{\begingroup\@sanitize@url \@url }%
\providecommand \@url [1]{\endgroup\@href {#1}{\urlprefix }}%
\providecommand \urlprefix  [0]{URL }%
\providecommand \Eprint [0]{\href }%
\providecommand \doibase [0]{https://doi.org/}%
\providecommand \selectlanguage [0]{\@gobble}%
\providecommand \bibinfo  [0]{\@secondoftwo}%
\providecommand \bibfield  [0]{\@secondoftwo}%
\providecommand \translation [1]{[#1]}%
\providecommand \BibitemOpen [0]{}%
\providecommand \bibitemStop [0]{}%
\providecommand \bibitemNoStop [0]{.\EOS\space}%
\providecommand \EOS [0]{\spacefactor3000\relax}%
\providecommand \BibitemShut  [1]{\csname bibitem#1\endcsname}%
\let\auto@bib@innerbib\@empty
\bibitem [{\citenamefont {Kempe}(2003)}]{Kempe2003}%
  \BibitemOpen
  \bibfield  {author} {\bibinfo {author} {\bibfnamefont {J.}~\bibnamefont
  {Kempe}},\ }\bibfield  {title} {\bibinfo {title} {Quantum random walks: An
  introductory overview},\ }\href@noop {} {\bibfield  {journal} {\bibinfo
  {journal} {Contemporary Physics}\ }\textbf {\bibinfo {volume} {44}},\
  \bibinfo {pages} {307} (\bibinfo {year} {2003})}\BibitemShut {NoStop}%
\bibitem [{\citenamefont {Portugal}(2018)}]{Portugal}%
  \BibitemOpen
  \bibfield  {author} {\bibinfo {author} {\bibfnamefont {R.}~\bibnamefont
  {Portugal}},\ }\href@noop {} {\emph {\bibinfo {title} {Quantum Walks and
  Search Algorithms}}}\ (\bibinfo  {publisher} {Springer International
  Publishing},\ \bibinfo {address} {New York},\ \bibinfo {year}
  {2018})\BibitemShut {NoStop}%
\bibitem [{\citenamefont {Lovett}\ \emph {et~al.}(2010)\citenamefont {Lovett},
  \citenamefont {Cooper}, \citenamefont {Everitt}, \citenamefont {Trevers},\
  and\ \citenamefont {Kendon}}]{Kendon}%
  \BibitemOpen
  \bibfield  {author} {\bibinfo {author} {\bibfnamefont {N.~B.}\ \bibnamefont
  {Lovett}}, \bibinfo {author} {\bibfnamefont {S.}~\bibnamefont {Cooper}},
  \bibinfo {author} {\bibfnamefont {M.}~\bibnamefont {Everitt}}, \bibinfo
  {author} {\bibfnamefont {M.}~\bibnamefont {Trevers}},\ and\ \bibinfo {author}
  {\bibfnamefont {V.}~\bibnamefont {Kendon}},\ }\bibfield  {title} {\bibinfo
  {title} {Universal quantum computation using the discrete-time quantum
  walk},\ }\href {https://doi.org/10.1103/PhysRevA.81.042330} {\bibfield
  {journal} {\bibinfo  {journal} {Phys. Rev. A}\ }\textbf {\bibinfo {volume}
  {81}},\ \bibinfo {pages} {042330} (\bibinfo {year} {2010})}\BibitemShut
  {NoStop}%
\bibitem [{\citenamefont {Weiss}(1994)}]{Weiss94}%
  \BibitemOpen
  \bibfield  {author} {\bibinfo {author} {\bibfnamefont {G.~H.}\ \bibnamefont
  {Weiss}},\ }\href@noop {} {\emph {\bibinfo {title} {Aspects and Applications
  of the Random Walk}}}\ (\bibinfo  {publisher} {North Holland},\ \bibinfo
  {address} {Amsterdam},\ \bibinfo {year} {1994})\BibitemShut {NoStop}%
\bibitem [{\citenamefont {Hern\'andez}\ and\ \citenamefont
  {Romanelli}(2013)}]{Romanelli2013}%
  \BibitemOpen
  \bibfield  {author} {\bibinfo {author} {\bibfnamefont {G.}~\bibnamefont
  {Hern\'andez}}\ and\ \bibinfo {author} {\bibfnamefont {A.}~\bibnamefont
  {Romanelli}},\ }\bibfield  {title} {\bibinfo {title} {Resonant quantum kicked
  rotor with two internal levels},\ }\href
  {https://doi.org/10.1103/PhysRevA.87.042316} {\bibfield  {journal} {\bibinfo
  {journal} {Phys. Rev. A}\ }\textbf {\bibinfo {volume} {87}},\ \bibinfo
  {pages} {042316} (\bibinfo {year} {2013})}\BibitemShut {NoStop}%
\bibitem [{\citenamefont {Clark}\ \emph {et~al.}(2021)\citenamefont {Clark},
  \citenamefont {Groiseau}, \citenamefont {Shaw}, \citenamefont {Dadras},
  \citenamefont {Binegar}, \citenamefont {Wimberger}, \citenamefont {Summy},\
  and\ \citenamefont {Liu}}]{Clark2021}%
  \BibitemOpen
  \bibfield  {author} {\bibinfo {author} {\bibfnamefont {J.~H.}\ \bibnamefont
  {Clark}}, \bibinfo {author} {\bibfnamefont {C.}~\bibnamefont {Groiseau}},
  \bibinfo {author} {\bibfnamefont {Z.~N.}\ \bibnamefont {Shaw}}, \bibinfo
  {author} {\bibfnamefont {S.}~\bibnamefont {Dadras}}, \bibinfo {author}
  {\bibfnamefont {C.}~\bibnamefont {Binegar}}, \bibinfo {author} {\bibfnamefont
  {S.}~\bibnamefont {Wimberger}}, \bibinfo {author} {\bibfnamefont {G.~S.}\
  \bibnamefont {Summy}},\ and\ \bibinfo {author} {\bibfnamefont
  {Y.}~\bibnamefont {Liu}},\ }\bibfield  {title} {\bibinfo {title} {Quantum to
  classical walk transitions tuned by spontaneous emissions},\ }\href
  {https://doi.org/10.1103/PhysRevResearch.3.043062} {\bibfield  {journal}
  {\bibinfo  {journal} {Phys. Rev. Research}\ }\textbf {\bibinfo {volume}
  {3}},\ \bibinfo {pages} {043062} (\bibinfo {year} {2021})}\BibitemShut
  {NoStop}%
\bibitem [{\citenamefont {Dadras}\ \emph {et~al.}(2019)\citenamefont {Dadras},
  \citenamefont {Gresch}, \citenamefont {Groiseau}, \citenamefont {Wimberger},\
  and\ \citenamefont {Summy}}]{dadras2019experimental}%
  \BibitemOpen
  \bibfield  {author} {\bibinfo {author} {\bibfnamefont {S.}~\bibnamefont
  {Dadras}}, \bibinfo {author} {\bibfnamefont {A.}~\bibnamefont {Gresch}},
  \bibinfo {author} {\bibfnamefont {C.}~\bibnamefont {Groiseau}}, \bibinfo
  {author} {\bibfnamefont {S.}~\bibnamefont {Wimberger}},\ and\ \bibinfo
  {author} {\bibfnamefont {G.~S.}\ \bibnamefont {Summy}},\ }\bibfield  {title}
  {\bibinfo {title} {Experimental realization of a momentum-space quantum
  walk},\ }\href@noop {} {\bibfield  {journal} {\bibinfo  {journal} {Physical
  Review A}\ }\textbf {\bibinfo {volume} {99}},\ \bibinfo {pages} {043617}
  (\bibinfo {year} {2019})}\BibitemShut {NoStop}%
\bibitem [{\citenamefont {Bolik}\ \emph {et~al.}(2022)\citenamefont {Bolik},
  \citenamefont {Groiseau}, \citenamefont {Clark}, \citenamefont {Gresch},
  \citenamefont {Dadras}, \citenamefont {Summy}, \citenamefont {Liu},\ and\
  \citenamefont {Wimberger}}]{Bolik2022}%
  \BibitemOpen
  \bibfield  {author} {\bibinfo {author} {\bibfnamefont {N.}~\bibnamefont
  {Bolik}}, \bibinfo {author} {\bibfnamefont {C.}~\bibnamefont {Groiseau}},
  \bibinfo {author} {\bibfnamefont {J.~H.}\ \bibnamefont {Clark}}, \bibinfo
  {author} {\bibfnamefont {A.}~\bibnamefont {Gresch}}, \bibinfo {author}
  {\bibfnamefont {S.}~\bibnamefont {Dadras}}, \bibinfo {author} {\bibfnamefont
  {G.~S.}\ \bibnamefont {Summy}}, \bibinfo {author} {\bibfnamefont
  {Y.}~\bibnamefont {Liu}},\ and\ \bibinfo {author} {\bibfnamefont
  {S.}~\bibnamefont {Wimberger}},\ }\bibfield  {title} {\bibinfo {title}
  {Light-shift-induced behaviors observed in momentum-space quantum walks},\
  }\href {https://doi.org/10.1103/PhysRevA.106.033307} {\bibfield  {journal}
  {\bibinfo  {journal} {Phys. Rev. A}\ }\textbf {\bibinfo {volume} {106}},\
  \bibinfo {pages} {033307} (\bibinfo {year} {2022})}\BibitemShut {NoStop}%
\bibitem [{\citenamefont {Dadras}\ \emph {et~al.}(2018)\citenamefont {Dadras},
  \citenamefont {Gresch}, \citenamefont {Groiseau}, \citenamefont {Wimberger},\
  and\ \citenamefont {Summy}}]{Dadras2018}%
  \BibitemOpen
  \bibfield  {author} {\bibinfo {author} {\bibfnamefont {S.}~\bibnamefont
  {Dadras}}, \bibinfo {author} {\bibfnamefont {A.}~\bibnamefont {Gresch}},
  \bibinfo {author} {\bibfnamefont {C.}~\bibnamefont {Groiseau}}, \bibinfo
  {author} {\bibfnamefont {S.}~\bibnamefont {Wimberger}},\ and\ \bibinfo
  {author} {\bibfnamefont {G.~S.}\ \bibnamefont {Summy}},\ }\bibfield  {title}
  {\bibinfo {title} {Quantum walk in momentum space with a {Bose-Einstein}
  condensate},\ }\href {https://doi.org/10.1103/PhysRevLett.121.070402}
  {\bibfield  {journal} {\bibinfo  {journal} {Phys. Rev. Lett.}\ }\textbf
  {\bibinfo {volume} {121}},\ \bibinfo {pages} {070402} (\bibinfo {year}
  {2018})}\BibitemShut {NoStop}%
\bibitem [{\citenamefont {Preiss}\ \emph {et~al.}(2015)\citenamefont {Preiss},
  \citenamefont {Ma}, \citenamefont {Tai}, \citenamefont {Lukin}, \citenamefont
  {Rispoli}, \citenamefont {Zupancic}, \citenamefont {Lahini}, \citenamefont
  {Islam},\ and\ \citenamefont {Greiner}}]{preiss2015strongly}%
  \BibitemOpen
  \bibfield  {author} {\bibinfo {author} {\bibfnamefont {P.~M.}\ \bibnamefont
  {Preiss}}, \bibinfo {author} {\bibfnamefont {R.}~\bibnamefont {Ma}}, \bibinfo
  {author} {\bibfnamefont {M.~E.}\ \bibnamefont {Tai}}, \bibinfo {author}
  {\bibfnamefont {A.}~\bibnamefont {Lukin}}, \bibinfo {author} {\bibfnamefont
  {M.}~\bibnamefont {Rispoli}}, \bibinfo {author} {\bibfnamefont
  {P.}~\bibnamefont {Zupancic}}, \bibinfo {author} {\bibfnamefont
  {Y.}~\bibnamefont {Lahini}}, \bibinfo {author} {\bibfnamefont
  {R.}~\bibnamefont {Islam}},\ and\ \bibinfo {author} {\bibfnamefont
  {M.}~\bibnamefont {Greiner}},\ }\bibfield  {title} {\bibinfo {title}
  {Strongly correlated quantum walks in optical lattices},\ }\href@noop {}
  {\bibfield  {journal} {\bibinfo  {journal} {Science}\ }\textbf {\bibinfo
  {volume} {347}},\ \bibinfo {pages} {1229} (\bibinfo {year}
  {2015})}\BibitemShut {NoStop}%
\bibitem [{\citenamefont {D{\"u}r}\ \emph {et~al.}(2002)\citenamefont
  {D{\"u}r}, \citenamefont {Raussendorf}, \citenamefont {Kendon},\ and\
  \citenamefont {Briegel}}]{dur2002quantum}%
  \BibitemOpen
  \bibfield  {author} {\bibinfo {author} {\bibfnamefont {W.}~\bibnamefont
  {D{\"u}r}}, \bibinfo {author} {\bibfnamefont {R.}~\bibnamefont
  {Raussendorf}}, \bibinfo {author} {\bibfnamefont {V.~M.}\ \bibnamefont
  {Kendon}},\ and\ \bibinfo {author} {\bibfnamefont {H.-J.}\ \bibnamefont
  {Briegel}},\ }\bibfield  {title} {\bibinfo {title} {Quantum walks in optical
  lattices},\ }\href@noop {} {\bibfield  {journal} {\bibinfo  {journal}
  {Physical Review A}\ }\textbf {\bibinfo {volume} {66}},\ \bibinfo {pages}
  {052319} (\bibinfo {year} {2002})}\BibitemShut {NoStop}%
\bibitem [{\citenamefont {Eckert}\ \emph {et~al.}(2005)\citenamefont {Eckert},
  \citenamefont {Mompart}, \citenamefont {Birkl},\ and\ \citenamefont
  {Lewenstein}}]{eckert2005one}%
  \BibitemOpen
  \bibfield  {author} {\bibinfo {author} {\bibfnamefont {K.}~\bibnamefont
  {Eckert}}, \bibinfo {author} {\bibfnamefont {J.}~\bibnamefont {Mompart}},
  \bibinfo {author} {\bibfnamefont {G.}~\bibnamefont {Birkl}},\ and\ \bibinfo
  {author} {\bibfnamefont {M.}~\bibnamefont {Lewenstein}},\ }\bibfield  {title}
  {\bibinfo {title} {One-and two-dimensional quantum walks in arrays of optical
  traps},\ }\href@noop {} {\bibfield  {journal} {\bibinfo  {journal} {Physical
  Review A}\ }\textbf {\bibinfo {volume} {72}},\ \bibinfo {pages} {012327}
  (\bibinfo {year} {2005})}\BibitemShut {NoStop}%
\bibitem [{\citenamefont {Steffen}\ \emph {et~al.}(2012)\citenamefont
  {Steffen}, \citenamefont {Alberti}, \citenamefont {Alt}, \citenamefont
  {Belmechri}, \citenamefont {Hild}, \citenamefont {Karski}, \citenamefont
  {Widera},\ and\ \citenamefont {Meschede}}]{steffen2012digital}%
  \BibitemOpen
  \bibfield  {author} {\bibinfo {author} {\bibfnamefont {A.}~\bibnamefont
  {Steffen}}, \bibinfo {author} {\bibfnamefont {A.}~\bibnamefont {Alberti}},
  \bibinfo {author} {\bibfnamefont {W.}~\bibnamefont {Alt}}, \bibinfo {author}
  {\bibfnamefont {N.}~\bibnamefont {Belmechri}}, \bibinfo {author}
  {\bibfnamefont {S.}~\bibnamefont {Hild}}, \bibinfo {author} {\bibfnamefont
  {M.}~\bibnamefont {Karski}}, \bibinfo {author} {\bibfnamefont
  {A.}~\bibnamefont {Widera}},\ and\ \bibinfo {author} {\bibfnamefont
  {D.}~\bibnamefont {Meschede}},\ }\bibfield  {title} {\bibinfo {title}
  {Digital atom interferometer with single particle control on a discretized
  space-time geometry},\ }\href@noop {} {\bibfield  {journal} {\bibinfo
  {journal} {Proceedings of the National Academy of Sciences}\ }\textbf
  {\bibinfo {volume} {109}},\ \bibinfo {pages} {9770} (\bibinfo {year}
  {2012})}\BibitemShut {NoStop}%
\bibitem [{\citenamefont {Groh}\ \emph {et~al.}(2016)\citenamefont {Groh},
  \citenamefont {Brakhane}, \citenamefont {Alt}, \citenamefont {Meschede},
  \citenamefont {Asb{\'o}th},\ and\ \citenamefont
  {Alberti}}]{groh2016robustness}%
  \BibitemOpen
  \bibfield  {author} {\bibinfo {author} {\bibfnamefont {T.}~\bibnamefont
  {Groh}}, \bibinfo {author} {\bibfnamefont {S.}~\bibnamefont {Brakhane}},
  \bibinfo {author} {\bibfnamefont {W.}~\bibnamefont {Alt}}, \bibinfo {author}
  {\bibfnamefont {D.}~\bibnamefont {Meschede}}, \bibinfo {author}
  {\bibfnamefont {J.~K.}\ \bibnamefont {Asb{\'o}th}},\ and\ \bibinfo {author}
  {\bibfnamefont {A.}~\bibnamefont {Alberti}},\ }\bibfield  {title} {\bibinfo
  {title} {Robustness of topologically protected edge states in quantum walk
  experiments with neutral atoms},\ }\href@noop {} {\bibfield  {journal}
  {\bibinfo  {journal} {Physical Review A}\ }\textbf {\bibinfo {volume} {94}},\
  \bibinfo {pages} {013620} (\bibinfo {year} {2016})}\BibitemShut {NoStop}%
\bibitem [{\citenamefont {Karski}\ \emph {et~al.}(2009)\citenamefont {Karski},
  \citenamefont {F{\"o}rster}, \citenamefont {Choi}, \citenamefont {Steffen},
  \citenamefont {Alt}, \citenamefont {Meschede},\ and\ \citenamefont
  {Widera}}]{karski2009quantum}%
  \BibitemOpen
  \bibfield  {author} {\bibinfo {author} {\bibfnamefont {M.}~\bibnamefont
  {Karski}}, \bibinfo {author} {\bibfnamefont {L.}~\bibnamefont {F{\"o}rster}},
  \bibinfo {author} {\bibfnamefont {J.-M.}\ \bibnamefont {Choi}}, \bibinfo
  {author} {\bibfnamefont {A.}~\bibnamefont {Steffen}}, \bibinfo {author}
  {\bibfnamefont {W.}~\bibnamefont {Alt}}, \bibinfo {author} {\bibfnamefont
  {D.}~\bibnamefont {Meschede}},\ and\ \bibinfo {author} {\bibfnamefont
  {A.}~\bibnamefont {Widera}},\ }\bibfield  {title} {\bibinfo {title} {Quantum
  walk in position space with single optically trapped atoms},\ }\href@noop {}
  {\bibfield  {journal} {\bibinfo  {journal} {Science}\ }\textbf {\bibinfo
  {volume} {325}},\ \bibinfo {pages} {174} (\bibinfo {year}
  {2009})}\BibitemShut {NoStop}%
\bibitem [{\citenamefont
  {Chandrashekar}(2006)}]{chandrashekar2006implementing}%
  \BibitemOpen
  \bibfield  {author} {\bibinfo {author} {\bibfnamefont {C.}~\bibnamefont
  {Chandrashekar}},\ }\bibfield  {title} {\bibinfo {title} {{Implementing the
  one-dimensional quantum (Hadamard) walk using a Bose-Einstein condensate}},\
  }\href@noop {} {\bibfield  {journal} {\bibinfo  {journal} {Physical Review
  A}\ }\textbf {\bibinfo {volume} {74}},\ \bibinfo {pages} {032307} (\bibinfo
  {year} {2006})}\BibitemShut {NoStop}%
\bibitem [{\citenamefont {Travaglione}\ and\ \citenamefont
  {Milburn}(2002)}]{travaglione2002implementing}%
  \BibitemOpen
  \bibfield  {author} {\bibinfo {author} {\bibfnamefont {B.~C.}\ \bibnamefont
  {Travaglione}}\ and\ \bibinfo {author} {\bibfnamefont {G.~J.}\ \bibnamefont
  {Milburn}},\ }\bibfield  {title} {\bibinfo {title} {Implementing the quantum
  random walk},\ }\href@noop {} {\bibfield  {journal} {\bibinfo  {journal}
  {Physical Review A}\ }\textbf {\bibinfo {volume} {65}},\ \bibinfo {pages}
  {032310} (\bibinfo {year} {2002})}\BibitemShut {NoStop}%
\bibitem [{\citenamefont {Z{\"a}hringer}\ \emph {et~al.}(2010)\citenamefont
  {Z{\"a}hringer}, \citenamefont {Kirchmair}, \citenamefont {Gerritsma},
  \citenamefont {Solano}, \citenamefont {Blatt},\ and\ \citenamefont
  {Roos}}]{zahringer2010realization}%
  \BibitemOpen
  \bibfield  {author} {\bibinfo {author} {\bibfnamefont {F.}~\bibnamefont
  {Z{\"a}hringer}}, \bibinfo {author} {\bibfnamefont {G.}~\bibnamefont
  {Kirchmair}}, \bibinfo {author} {\bibfnamefont {R.}~\bibnamefont
  {Gerritsma}}, \bibinfo {author} {\bibfnamefont {E.}~\bibnamefont {Solano}},
  \bibinfo {author} {\bibfnamefont {R.}~\bibnamefont {Blatt}},\ and\ \bibinfo
  {author} {\bibfnamefont {C.}~\bibnamefont {Roos}},\ }\bibfield  {title}
  {\bibinfo {title} {Realization of a quantum walk with one and two trapped
  ions},\ }\href@noop {} {\bibfield  {journal} {\bibinfo  {journal} {Phys. Rev.
  Lett.}\ }\textbf {\bibinfo {volume} {104}},\ \bibinfo {pages} {100503}
  (\bibinfo {year} {2010})}\BibitemShut {NoStop}%
\bibitem [{\citenamefont {Schmitz}\ \emph {et~al.}(2009)\citenamefont
  {Schmitz}, \citenamefont {Matjeschk}, \citenamefont {Schneider},
  \citenamefont {Glueckert}, \citenamefont {Enderlein}, \citenamefont {Huber},\
  and\ \citenamefont {Schaetz}}]{schmitz2009quantum}%
  \BibitemOpen
  \bibfield  {author} {\bibinfo {author} {\bibfnamefont {H.}~\bibnamefont
  {Schmitz}}, \bibinfo {author} {\bibfnamefont {R.}~\bibnamefont {Matjeschk}},
  \bibinfo {author} {\bibfnamefont {C.}~\bibnamefont {Schneider}}, \bibinfo
  {author} {\bibfnamefont {J.}~\bibnamefont {Glueckert}}, \bibinfo {author}
  {\bibfnamefont {M.}~\bibnamefont {Enderlein}}, \bibinfo {author}
  {\bibfnamefont {T.}~\bibnamefont {Huber}},\ and\ \bibinfo {author}
  {\bibfnamefont {T.}~\bibnamefont {Schaetz}},\ }\bibfield  {title} {\bibinfo
  {title} {Quantum walk of a trapped ion in phase space},\ }\href@noop {}
  {\bibfield  {journal} {\bibinfo  {journal} {Phys. Rev. Lett.}\ }\textbf
  {\bibinfo {volume} {103}},\ \bibinfo {pages} {090504} (\bibinfo {year}
  {2009})}\BibitemShut {NoStop}%
\bibitem [{\citenamefont {Perets}\ \emph {et~al.}(2008)\citenamefont {Perets},
  \citenamefont {Lahini}, \citenamefont {Pozzi}, \citenamefont {Sorel},
  \citenamefont {Morandotti},\ and\ \citenamefont
  {Silberberg}}]{perets2008realization}%
  \BibitemOpen
  \bibfield  {author} {\bibinfo {author} {\bibfnamefont {H.~B.}\ \bibnamefont
  {Perets}}, \bibinfo {author} {\bibfnamefont {Y.}~\bibnamefont {Lahini}},
  \bibinfo {author} {\bibfnamefont {F.}~\bibnamefont {Pozzi}}, \bibinfo
  {author} {\bibfnamefont {M.}~\bibnamefont {Sorel}}, \bibinfo {author}
  {\bibfnamefont {R.}~\bibnamefont {Morandotti}},\ and\ \bibinfo {author}
  {\bibfnamefont {Y.}~\bibnamefont {Silberberg}},\ }\bibfield  {title}
  {\bibinfo {title} {Realization of quantum walks with negligible decoherence
  in waveguide lattices},\ }\href@noop {} {\bibfield  {journal} {\bibinfo
  {journal} {Phys. Rev. Lett.}\ }\textbf {\bibinfo {volume} {100}},\ \bibinfo
  {pages} {170506} (\bibinfo {year} {2008})}\BibitemShut {NoStop}%
\bibitem [{\citenamefont {Peruzzo}\ \emph {et~al.}(2010)\citenamefont
  {Peruzzo}, \citenamefont {Lobino}, \citenamefont {Matthews}, \citenamefont
  {Matsuda}, \citenamefont {Politi}, \citenamefont {Poulios}, \citenamefont
  {Zhou}, \citenamefont {Lahini}, \citenamefont {Ismail}, \citenamefont
  {W{\"o}rhoff} \emph {et~al.}}]{peruzzo2010quantum}%
  \BibitemOpen
  \bibfield  {author} {\bibinfo {author} {\bibfnamefont {A.}~\bibnamefont
  {Peruzzo}}, \bibinfo {author} {\bibfnamefont {M.}~\bibnamefont {Lobino}},
  \bibinfo {author} {\bibfnamefont {J.~C.}\ \bibnamefont {Matthews}}, \bibinfo
  {author} {\bibfnamefont {N.}~\bibnamefont {Matsuda}}, \bibinfo {author}
  {\bibfnamefont {A.}~\bibnamefont {Politi}}, \bibinfo {author} {\bibfnamefont
  {K.}~\bibnamefont {Poulios}}, \bibinfo {author} {\bibfnamefont {X.-Q.}\
  \bibnamefont {Zhou}}, \bibinfo {author} {\bibfnamefont {Y.}~\bibnamefont
  {Lahini}}, \bibinfo {author} {\bibfnamefont {N.}~\bibnamefont {Ismail}},
  \bibinfo {author} {\bibfnamefont {K.}~\bibnamefont {W{\"o}rhoff}}, \emph
  {et~al.},\ }\bibfield  {title} {\bibinfo {title} {Quantum walks of correlated
  photons},\ }\href@noop {} {\bibfield  {journal} {\bibinfo  {journal}
  {Science}\ }\textbf {\bibinfo {volume} {329}},\ \bibinfo {pages} {1500}
  (\bibinfo {year} {2010})}\BibitemShut {NoStop}%
\bibitem [{\citenamefont {Cardano}\ \emph {et~al.}(2017)\citenamefont
  {Cardano}, \citenamefont {D'Errico}, \citenamefont {Dauphin}, \citenamefont
  {Maffei}, \citenamefont {Piccirillo}, \citenamefont {de~Lisio}, \citenamefont
  {De~Filippis}, \citenamefont {Cataudella}, \citenamefont {Santamato},
  \citenamefont {Marrucci} \emph {et~al.}}]{cardano2017detection}%
  \BibitemOpen
  \bibfield  {author} {\bibinfo {author} {\bibfnamefont {F.}~\bibnamefont
  {Cardano}}, \bibinfo {author} {\bibfnamefont {A.}~\bibnamefont {D'Errico}},
  \bibinfo {author} {\bibfnamefont {A.}~\bibnamefont {Dauphin}}, \bibinfo
  {author} {\bibfnamefont {M.}~\bibnamefont {Maffei}}, \bibinfo {author}
  {\bibfnamefont {B.}~\bibnamefont {Piccirillo}}, \bibinfo {author}
  {\bibfnamefont {C.}~\bibnamefont {de~Lisio}}, \bibinfo {author}
  {\bibfnamefont {G.}~\bibnamefont {De~Filippis}}, \bibinfo {author}
  {\bibfnamefont {V.}~\bibnamefont {Cataudella}}, \bibinfo {author}
  {\bibfnamefont {E.}~\bibnamefont {Santamato}}, \bibinfo {author}
  {\bibfnamefont {L.}~\bibnamefont {Marrucci}}, \emph {et~al.},\ }\bibfield
  {title} {\bibinfo {title} {Detection of {Zak} phases and topological
  invariants in a chiral quantum walk of twisted photons},\ }\href@noop {}
  {\bibfield  {journal} {\bibinfo  {journal} {Nature Communications}\ }\textbf
  {\bibinfo {volume} {8}},\ \bibinfo {pages} {15516} (\bibinfo {year}
  {2017})}\BibitemShut {NoStop}%
\bibitem [{\citenamefont {Chen}\ \emph {et~al.}(2018)\citenamefont {Chen},
  \citenamefont {Ding}, \citenamefont {Qin}, \citenamefont {He}, \citenamefont
  {Luo}, \citenamefont {Chen}, \citenamefont {Liu}, \citenamefont {Wang},
  \citenamefont {Zhang}, \citenamefont {Li} \emph
  {et~al.}}]{chen2018observation}%
  \BibitemOpen
  \bibfield  {author} {\bibinfo {author} {\bibfnamefont {C.}~\bibnamefont
  {Chen}}, \bibinfo {author} {\bibfnamefont {X.}~\bibnamefont {Ding}}, \bibinfo
  {author} {\bibfnamefont {J.}~\bibnamefont {Qin}}, \bibinfo {author}
  {\bibfnamefont {Y.}~\bibnamefont {He}}, \bibinfo {author} {\bibfnamefont
  {Y.-H.}\ \bibnamefont {Luo}}, \bibinfo {author} {\bibfnamefont {M.-C.}\
  \bibnamefont {Chen}}, \bibinfo {author} {\bibfnamefont {C.}~\bibnamefont
  {Liu}}, \bibinfo {author} {\bibfnamefont {X.-L.}\ \bibnamefont {Wang}},
  \bibinfo {author} {\bibfnamefont {W.-J.}\ \bibnamefont {Zhang}}, \bibinfo
  {author} {\bibfnamefont {H.}~\bibnamefont {Li}}, \emph {et~al.},\ }\bibfield
  {title} {\bibinfo {title} {Observation of topologically protected edge states
  in a photonic two-dimensional quantum walk},\ }\href@noop {} {\bibfield
  {journal} {\bibinfo  {journal} {Phys. Rev. Lett.}\ }\textbf {\bibinfo
  {volume} {121}},\ \bibinfo {pages} {100502} (\bibinfo {year}
  {2018})}\BibitemShut {NoStop}%
\bibitem [{\citenamefont {Tang}\ \emph {et~al.}(2018)\citenamefont {Tang},
  \citenamefont {Lin}, \citenamefont {Feng}, \citenamefont {Chen},
  \citenamefont {Gao}, \citenamefont {Sun}, \citenamefont {Wang}, \citenamefont
  {Lai}, \citenamefont {Xu}, \citenamefont {Wang} \emph
  {et~al.}}]{tang2018experimental}%
  \BibitemOpen
  \bibfield  {author} {\bibinfo {author} {\bibfnamefont {H.}~\bibnamefont
  {Tang}}, \bibinfo {author} {\bibfnamefont {X.-F.}\ \bibnamefont {Lin}},
  \bibinfo {author} {\bibfnamefont {Z.}~\bibnamefont {Feng}}, \bibinfo {author}
  {\bibfnamefont {J.-Y.}\ \bibnamefont {Chen}}, \bibinfo {author}
  {\bibfnamefont {J.}~\bibnamefont {Gao}}, \bibinfo {author} {\bibfnamefont
  {K.}~\bibnamefont {Sun}}, \bibinfo {author} {\bibfnamefont {C.-Y.}\
  \bibnamefont {Wang}}, \bibinfo {author} {\bibfnamefont {P.-C.}\ \bibnamefont
  {Lai}}, \bibinfo {author} {\bibfnamefont {X.-Y.}\ \bibnamefont {Xu}},
  \bibinfo {author} {\bibfnamefont {Y.}~\bibnamefont {Wang}}, \emph {et~al.},\
  }\bibfield  {title} {\bibinfo {title} {Experimental two-dimensional quantum
  walk on a photonic chip},\ }\href@noop {} {\bibfield  {journal} {\bibinfo
  {journal} {Sci. Adv.}\ }\textbf {\bibinfo {volume} {4}},\ \bibinfo {pages}
  {eaat3174} (\bibinfo {year} {2018})}\BibitemShut {NoStop}%
\bibitem [{\citenamefont {Poulios}\ \emph {et~al.}(2014)\citenamefont
  {Poulios}, \citenamefont {Keil}, \citenamefont {Fry}, \citenamefont
  {Meinecke}, \citenamefont {Matthews}, \citenamefont {Politi}, \citenamefont
  {Lobino}, \citenamefont {Gr{\"a}fe}, \citenamefont {Heinrich}, \citenamefont
  {Nolte} \emph {et~al.}}]{poulios2014quantum}%
  \BibitemOpen
  \bibfield  {author} {\bibinfo {author} {\bibfnamefont {K.}~\bibnamefont
  {Poulios}}, \bibinfo {author} {\bibfnamefont {R.}~\bibnamefont {Keil}},
  \bibinfo {author} {\bibfnamefont {D.}~\bibnamefont {Fry}}, \bibinfo {author}
  {\bibfnamefont {J.~D.}\ \bibnamefont {Meinecke}}, \bibinfo {author}
  {\bibfnamefont {J.~C.}\ \bibnamefont {Matthews}}, \bibinfo {author}
  {\bibfnamefont {A.}~\bibnamefont {Politi}}, \bibinfo {author} {\bibfnamefont
  {M.}~\bibnamefont {Lobino}}, \bibinfo {author} {\bibfnamefont
  {M.}~\bibnamefont {Gr{\"a}fe}}, \bibinfo {author} {\bibfnamefont
  {M.}~\bibnamefont {Heinrich}}, \bibinfo {author} {\bibfnamefont
  {S.}~\bibnamefont {Nolte}}, \emph {et~al.},\ }\bibfield  {title} {\bibinfo
  {title} {Quantum walks of correlated photon pairs in two-dimensional
  waveguide arrays},\ }\href@noop {} {\bibfield  {journal} {\bibinfo  {journal}
  {Phys. Rev. Lett.}\ }\textbf {\bibinfo {volume} {112}},\ \bibinfo {pages}
  {143604} (\bibinfo {year} {2014})}\BibitemShut {NoStop}%
\bibitem [{\citenamefont {Schreiber}\ \emph {et~al.}(2010)\citenamefont
  {Schreiber}, \citenamefont {Cassemiro}, \citenamefont {Poto{\v{c}}ek},
  \citenamefont {G{\'a}bris}, \citenamefont {Mosley}, \citenamefont
  {Andersson}, \citenamefont {Jex},\ and\ \citenamefont
  {Silberhorn}}]{schreiber2010photons}%
  \BibitemOpen
  \bibfield  {author} {\bibinfo {author} {\bibfnamefont {A.}~\bibnamefont
  {Schreiber}}, \bibinfo {author} {\bibfnamefont {K.~N.}\ \bibnamefont
  {Cassemiro}}, \bibinfo {author} {\bibfnamefont {V.}~\bibnamefont
  {Poto{\v{c}}ek}}, \bibinfo {author} {\bibfnamefont {A.}~\bibnamefont
  {G{\'a}bris}}, \bibinfo {author} {\bibfnamefont {P.~J.}\ \bibnamefont
  {Mosley}}, \bibinfo {author} {\bibfnamefont {E.}~\bibnamefont {Andersson}},
  \bibinfo {author} {\bibfnamefont {I.}~\bibnamefont {Jex}},\ and\ \bibinfo
  {author} {\bibfnamefont {C.}~\bibnamefont {Silberhorn}},\ }\bibfield  {title}
  {\bibinfo {title} {Photons walking the line: a quantum walk with adjustable
  coin operations},\ }\href@noop {} {\bibfield  {journal} {\bibinfo  {journal}
  {Phys. Rev. Lett.}\ }\textbf {\bibinfo {volume} {104}},\ \bibinfo {pages}
  {050502} (\bibinfo {year} {2010})}\BibitemShut {NoStop}%
\bibitem [{\citenamefont {Raizen}(1999)}]{Raizen1999}%
  \BibitemOpen
  \bibfield  {author} {\bibinfo {author} {\bibfnamefont {M.~G.}\ \bibnamefont
  {Raizen}},\ }\bibfield  {title} {\bibinfo {title} {Quantum chaos with cold
  atoms},\ }\href@noop {} {\bibfield  {journal} {\bibinfo  {journal} {Advances
  in Atomic, Molecular, and Optical Physics}\ }\textbf {\bibinfo {volume}
  {41}},\ \bibinfo {pages} {199} (\bibinfo {year} {1999})}\BibitemShut
  {NoStop}%
\bibitem [{\citenamefont {Sadgrove}\ and\ \citenamefont
  {Wimberger}(2011)}]{SW2011}%
  \BibitemOpen
  \bibfield  {author} {\bibinfo {author} {\bibfnamefont {M.}~\bibnamefont
  {Sadgrove}}\ and\ \bibinfo {author} {\bibfnamefont {S.}~\bibnamefont
  {Wimberger}},\ }\bibfield  {title} {\bibinfo {title} {{A pseudo-classical
  method for the atom-optics kicked rotor: from theory to experiment and
  back}},\ }\href@noop {} {\bibfield  {journal} {\bibinfo  {journal} {Adv. At.
  Mol. Opt. Phys.}\ }\textbf {\bibinfo {volume} {60}},\ \bibinfo {pages} {315}
  (\bibinfo {year} {2011})}\BibitemShut {NoStop}%
\bibitem [{\citenamefont {Santhanam}\ \emph {et~al.}(2022)\citenamefont
  {Santhanam}, \citenamefont {Paul},\ and\ \citenamefont {Kannan}}]{Rev2022}%
  \BibitemOpen
  \bibfield  {author} {\bibinfo {author} {\bibfnamefont {M.}~\bibnamefont
  {Santhanam}}, \bibinfo {author} {\bibfnamefont {S.}~\bibnamefont {Paul}},\
  and\ \bibinfo {author} {\bibfnamefont {J.~B.}\ \bibnamefont {Kannan}},\
  }\bibfield  {title} {\bibinfo {title} {Quantum kicked rotor and its variants:
  Chaos, localization and beyond},\ }\href
  {https://doi.org/https://doi.org/10.1016/j.physrep.2022.01.002} {\bibfield
  {journal} {\bibinfo  {journal} {Phys. Rep.}\ }\textbf {\bibinfo {volume}
  {956}},\ \bibinfo {pages} {1} (\bibinfo {year} {2022})}\BibitemShut {NoStop}%
\bibitem [{\citenamefont {Summy}\ and\ \citenamefont
  {Wimberger}(2016)}]{Summy2016}%
  \BibitemOpen
  \bibfield  {author} {\bibinfo {author} {\bibfnamefont {G.}~\bibnamefont
  {Summy}}\ and\ \bibinfo {author} {\bibfnamefont {S.}~\bibnamefont
  {Wimberger}},\ }\bibfield  {title} {\bibinfo {title} {{Quantum random walk of
  a Bose-Einstein condensate in momentum space}},\ }\href
  {https://doi.org/10.1103/PhysRevA.93.023638} {\bibfield  {journal} {\bibinfo
  {journal} {Phys. Rev. A}\ }\textbf {\bibinfo {volume} {93}},\ \bibinfo
  {pages} {023638} (\bibinfo {year} {2016})}\BibitemShut {NoStop}%
\bibitem [{\citenamefont {Wimberger}\ \emph {et~al.}(2003)\citenamefont
  {Wimberger}, \citenamefont {Guarneri},\ and\ \citenamefont
  {Fishman}}]{SandroWimberger_2003}%
  \BibitemOpen
  \bibfield  {author} {\bibinfo {author} {\bibfnamefont {S.}~\bibnamefont
  {Wimberger}}, \bibinfo {author} {\bibfnamefont {I.}~\bibnamefont
  {Guarneri}},\ and\ \bibinfo {author} {\bibfnamefont {S.}~\bibnamefont
  {Fishman}},\ }\bibfield  {title} {\bibinfo {title} {Quantum resonances and
  decoherence for $\delta$-kicked atoms},\ }\href
  {https://doi.org/10.1088/0951-7715/16/4/312} {\bibfield  {journal} {\bibinfo
  {journal} {Nonlinearity}\ }\textbf {\bibinfo {volume} {16}},\ \bibinfo
  {pages} {1381} (\bibinfo {year} {2003})}\BibitemShut {NoStop}%
\bibitem [{\citenamefont {Delvecchio}\ \emph {et~al.}(2020)\citenamefont
  {Delvecchio}, \citenamefont {Petiziol},\ and\ \citenamefont
  {Wimberger}}]{Delvecchio2020}%
  \BibitemOpen
  \bibfield  {author} {\bibinfo {author} {\bibfnamefont {M.}~\bibnamefont
  {Delvecchio}}, \bibinfo {author} {\bibfnamefont {F.}~\bibnamefont
  {Petiziol}},\ and\ \bibinfo {author} {\bibfnamefont {S.}~\bibnamefont
  {Wimberger}},\ }\bibfield  {title} {\bibinfo {title} {Resonant quantum kicked
  rotor as a continuous-time quantum walk},\ }\href
  {https://doi.org/10.3390/condmat5010004} {\bibfield  {journal} {\bibinfo
  {journal} {Condensed Matter}\ }\textbf {\bibinfo {volume} {5}},\ \bibinfo
  {pages} {4} (\bibinfo {year} {2020})}\BibitemShut {NoStop}%
\bibitem [{\citenamefont {Ni}\ \emph {et~al.}(2016)\citenamefont {Ni},
  \citenamefont {Lam}, \citenamefont {Dadras}, \citenamefont {Borunda},
  \citenamefont {Wimberger},\ and\ \citenamefont {Summy}}]{Ni2016}%
  \BibitemOpen
  \bibfield  {author} {\bibinfo {author} {\bibfnamefont {J.}~\bibnamefont
  {Ni}}, \bibinfo {author} {\bibfnamefont {W.~K.}\ \bibnamefont {Lam}},
  \bibinfo {author} {\bibfnamefont {S.}~\bibnamefont {Dadras}}, \bibinfo
  {author} {\bibfnamefont {M.~F.}\ \bibnamefont {Borunda}}, \bibinfo {author}
  {\bibfnamefont {S.}~\bibnamefont {Wimberger}},\ and\ \bibinfo {author}
  {\bibfnamefont {G.~S.}\ \bibnamefont {Summy}},\ }\bibfield  {title} {\bibinfo
  {title} {Initial-state dependence of a quantum resonance ratchet},\ }\href
  {https://doi.org/10.1103/PhysRevA.94.043620} {\bibfield  {journal} {\bibinfo
  {journal} {Phys. Rev. A}\ }\textbf {\bibinfo {volume} {94}},\ \bibinfo
  {pages} {043620} (\bibinfo {year} {2016})}\BibitemShut {NoStop}%
\bibitem [{\citenamefont {Schlunk}\ \emph {et~al.}(2003)\citenamefont
  {Schlunk}, \citenamefont {d'Arcy}, \citenamefont {Gardiner}, \citenamefont
  {Cassettari}, \citenamefont {Godun},\ and\ \citenamefont
  {Summy}}]{PhysRevLett.90.054101}%
  \BibitemOpen
  \bibfield  {author} {\bibinfo {author} {\bibfnamefont {S.}~\bibnamefont
  {Schlunk}}, \bibinfo {author} {\bibfnamefont {M.~B.}\ \bibnamefont {d'Arcy}},
  \bibinfo {author} {\bibfnamefont {S.~A.}\ \bibnamefont {Gardiner}}, \bibinfo
  {author} {\bibfnamefont {D.}~\bibnamefont {Cassettari}}, \bibinfo {author}
  {\bibfnamefont {R.~M.}\ \bibnamefont {Godun}},\ and\ \bibinfo {author}
  {\bibfnamefont {G.~S.}\ \bibnamefont {Summy}},\ }\bibfield  {title} {\bibinfo
  {title} {Signatures of quantum stability in a classically chaotic system},\
  }\href {https://doi.org/10.1103/PhysRevLett.90.054101} {\bibfield  {journal}
  {\bibinfo  {journal} {Phys. Rev. Lett.}\ }\textbf {\bibinfo {volume} {90}},\
  \bibinfo {pages} {054101} (\bibinfo {year} {2003})}\BibitemShut {NoStop}%
\bibitem [{\citenamefont {Chakraborty}\ \emph {et~al.}(2017)\citenamefont
  {Chakraborty}, \citenamefont {Das}, \citenamefont {Mallick},\ and\
  \citenamefont {Chandrashekar}}]{Chakraborty_2017}%
  \BibitemOpen
  \bibfield  {author} {\bibinfo {author} {\bibfnamefont {S.}~\bibnamefont
  {Chakraborty}}, \bibinfo {author} {\bibfnamefont {A.}~\bibnamefont {Das}},
  \bibinfo {author} {\bibfnamefont {A.}~\bibnamefont {Mallick}},\ and\ \bibinfo
  {author} {\bibfnamefont {C.~M.}\ \bibnamefont {Chandrashekar}},\ }\bibfield
  {title} {\bibinfo {title} {Quantum ratchet in disordered quantum walk},\
  }\href {https://doi.org/10.1002/andp.201600346} {\bibfield  {journal}
  {\bibinfo  {journal} {Annalen der Physik}\ }\textbf {\bibinfo {volume}
  {529}},\ \bibinfo {pages} {1600346} (\bibinfo {year} {2017})}\BibitemShut
  {NoStop}%
\bibitem [{\citenamefont {Trautmann}\ \emph {et~al.}(2022)\citenamefont
  {Trautmann}, \citenamefont {Groiseau},\ and\ \citenamefont
  {Wimberger}}]{Trautmann_2022}%
  \BibitemOpen
  \bibfield  {author} {\bibinfo {author} {\bibfnamefont {G.}~\bibnamefont
  {Trautmann}}, \bibinfo {author} {\bibfnamefont {C.}~\bibnamefont
  {Groiseau}},\ and\ \bibinfo {author} {\bibfnamefont {S.}~\bibnamefont
  {Wimberger}},\ }\bibfield  {title} {\bibinfo {title} {Parrondo's paradox for
  discrete-time quantum walks in momentum space},\ }\href
  {https://doi.org/10.1142/s0219477522500535} {\bibfield  {journal} {\bibinfo
  {journal} {Fluctuation and Noise Letters}\ }\textbf {\bibinfo {volume}
  {21}},\ \bibinfo {pages} {2250053} (\bibinfo {year} {2022})}\BibitemShut
  {NoStop}%
\bibitem [{\citenamefont {Chai}\ and\ \citenamefont
  {Andersen}(2020)}]{Andersen2020}%
  \BibitemOpen
  \bibfield  {author} {\bibinfo {author} {\bibfnamefont {S.}~\bibnamefont
  {Chai}}\ and\ \bibinfo {author} {\bibfnamefont {M.~F.}\ \bibnamefont
  {Andersen}},\ }\bibfield  {title} {\bibinfo {title} {Enhancing survival
  resonances with engineered dissipation},\ }\href
  {https://doi.org/10.1103/PhysRevResearch.2.033194} {\bibfield  {journal}
  {\bibinfo  {journal} {Phys. Rev. Res.}\ }\textbf {\bibinfo {volume} {2}},\
  \bibinfo {pages} {033194} (\bibinfo {year} {2020})}\BibitemShut {NoStop}%
\bibitem [{\citenamefont {Andersen}\ and\ \citenamefont
  {Wimberger}(2022)}]{Andersen2022}%
  \BibitemOpen
  \bibfield  {author} {\bibinfo {author} {\bibfnamefont {M.~F.}\ \bibnamefont
  {Andersen}}\ and\ \bibinfo {author} {\bibfnamefont {S.}~\bibnamefont
  {Wimberger}},\ }\bibfield  {title} {\bibinfo {title} {Classical model for
  survival resonances close to the talbot time},\ }\href
  {https://doi.org/10.1103/PhysRevA.105.013322} {\bibfield  {journal} {\bibinfo
   {journal} {Phys. Rev. A}\ }\textbf {\bibinfo {volume} {105}},\ \bibinfo
  {pages} {013322} (\bibinfo {year} {2022})}\BibitemShut {NoStop}%
\bibitem [{\citenamefont {Groiseau}(2017)}]{groiseau2017discrete}%
  \BibitemOpen
  \bibfield  {author} {\bibinfo {author} {\bibfnamefont {C.}~\bibnamefont
  {Groiseau}},\ }\emph {\bibinfo {title} {Discrete-time quantum walks in
  momentum space}},\ \href@noop {} {Master's thesis},\ \bibinfo  {school}
  {Heidelberg University} (\bibinfo {year} {2017})\BibitemShut {NoStop}%
\bibitem [{\citenamefont {Groiseau}\ and\ \citenamefont
  {Wimberger}(2019)}]{Caspar2019}%
  \BibitemOpen
  \bibfield  {author} {\bibinfo {author} {\bibfnamefont {C.}~\bibnamefont
  {Groiseau}}\ and\ \bibinfo {author} {\bibfnamefont {S.}~\bibnamefont
  {Wimberger}},\ }\bibfield  {title} {\bibinfo {title} {{Spontaneous emission
  in quantum walks of a kicked Bose-Einstein condensate}},\ }\href
  {https://doi.org/10.1103/PhysRevA.99.013610} {\bibfield  {journal} {\bibinfo
  {journal} {Phys. Rev. A}\ }\textbf {\bibinfo {volume} {99}},\ \bibinfo
  {pages} {013610} (\bibinfo {year} {2019})}\BibitemShut {NoStop}%
\bibitem [{\citenamefont {Wimberger}(2016)}]{Wimberger2016}%
  \BibitemOpen
  \bibfield  {author} {\bibinfo {author} {\bibfnamefont {S.}~\bibnamefont
  {Wimberger}},\ }\bibfield  {title} {\bibinfo {title} {Applications of
  fidelity measures to complex quantum systems},\ }\href@noop {} {\bibfield
  {journal} {\bibinfo  {journal} {Phil. Trans. Royal Soc. A (London)}\ }\textbf
  {\bibinfo {volume} {374}},\ \bibinfo {pages} {20150153} (\bibinfo {year}
  {2016})}\BibitemShut {NoStop}%
\bibitem [{\citenamefont {Ryu}\ \emph {et~al.}(2006)\citenamefont {Ryu},
  \citenamefont {Andersen}, \citenamefont {Vaziri}, \citenamefont {d'Arcy},
  \citenamefont {Grossman}, \citenamefont {Helmerson},\ and\ \citenamefont
  {Phillips}}]{Ryu2006}%
  \BibitemOpen
  \bibfield  {author} {\bibinfo {author} {\bibfnamefont {C.}~\bibnamefont
  {Ryu}}, \bibinfo {author} {\bibfnamefont {M.~F.}\ \bibnamefont {Andersen}},
  \bibinfo {author} {\bibfnamefont {A.}~\bibnamefont {Vaziri}}, \bibinfo
  {author} {\bibfnamefont {M.~B.}\ \bibnamefont {d'Arcy}}, \bibinfo {author}
  {\bibfnamefont {J.~M.}\ \bibnamefont {Grossman}}, \bibinfo {author}
  {\bibfnamefont {K.}~\bibnamefont {Helmerson}},\ and\ \bibinfo {author}
  {\bibfnamefont {W.~D.}\ \bibnamefont {Phillips}},\ }\bibfield  {title}
  {\bibinfo {title} {High-order quantum resonances observed in a periodically
  kicked bose-einstein condensate},\ }\href
  {https://doi.org/10.1103/PhysRevLett.96.160403} {\bibfield  {journal}
  {\bibinfo  {journal} {Phys. Rev. Lett.}\ }\textbf {\bibinfo {volume} {96}},\
  \bibinfo {pages} {160403} (\bibinfo {year} {2006})}\BibitemShut {NoStop}%
\bibitem [{\citenamefont {Liu}(2022)}]{Yingmei}%
  \BibitemOpen
  \bibfield  {author} {\bibinfo {author} {\bibfnamefont {Y.}~\bibnamefont
  {Liu}},\ }\href@noop {} {} (\bibinfo {year} {2022}),\ \bibinfo {note}
  {private communication}\BibitemShut {NoStop}%
\bibitem [{\citenamefont {Moore}\ \emph {et~al.}(1995)\citenamefont {Moore},
  \citenamefont {Robinson}, \citenamefont {Bharucha}, \citenamefont
  {Sundaram},\ and\ \citenamefont {Raizen}}]{PhysRevLett.75.4598}%
  \BibitemOpen
  \bibfield  {author} {\bibinfo {author} {\bibfnamefont {F.~L.}\ \bibnamefont
  {Moore}}, \bibinfo {author} {\bibfnamefont {J.~C.}\ \bibnamefont {Robinson}},
  \bibinfo {author} {\bibfnamefont {C.~F.}\ \bibnamefont {Bharucha}}, \bibinfo
  {author} {\bibfnamefont {B.}~\bibnamefont {Sundaram}},\ and\ \bibinfo
  {author} {\bibfnamefont {M.~G.}\ \bibnamefont {Raizen}},\ }\bibfield  {title}
  {\bibinfo {title} {Atom optics realization of the quantum
  $\ensuremath{\delta}$-kicked rotor},\ }\href
  {https://doi.org/10.1103/PhysRevLett.75.4598} {\bibfield  {journal} {\bibinfo
   {journal} {Phys. Rev. Lett.}\ }\textbf {\bibinfo {volume} {75}},\ \bibinfo
  {pages} {4598} (\bibinfo {year} {1995})}\BibitemShut {NoStop}%
\bibitem [{\citenamefont {Klappauf}\ \emph {et~al.}(1999)\citenamefont
  {Klappauf}, \citenamefont {Oskay}, \citenamefont {Steck},\ and\ \citenamefont
  {Raizen}}]{Klapp1999}%
  \BibitemOpen
  \bibfield  {author} {\bibinfo {author} {\bibfnamefont {B.}~\bibnamefont
  {Klappauf}}, \bibinfo {author} {\bibfnamefont {W.}~\bibnamefont {Oskay}},
  \bibinfo {author} {\bibfnamefont {D.}~\bibnamefont {Steck}},\ and\ \bibinfo
  {author} {\bibfnamefont {M.}~\bibnamefont {Raizen}},\ }\bibfield  {title}
  {\bibinfo {title} {Quantum chaos with cesium atoms: pushing the boundaries},\
  }\href {https://doi.org/https://doi.org/10.1016/S0167-2789(98)00221-8}
  {\bibfield  {journal} {\bibinfo  {journal} {Physica D}\ }\textbf {\bibinfo
  {volume} {131}},\ \bibinfo {pages} {78} (\bibinfo {year} {1999})}\BibitemShut
  {NoStop}%
\bibitem [{\citenamefont {Bl\"umel}\ \emph {et~al.}(1986)\citenamefont
  {Bl\"umel}, \citenamefont {Fishman},\ and\ \citenamefont
  {Smilansky}}]{Fishman1986}%
  \BibitemOpen
  \bibfield  {author} {\bibinfo {author} {\bibfnamefont {R.}~\bibnamefont
  {Bl\"umel}}, \bibinfo {author} {\bibfnamefont {S.}~\bibnamefont {Fishman}},\
  and\ \bibinfo {author} {\bibfnamefont {U.}~\bibnamefont {Smilansky}},\
  }\bibfield  {title} {\bibinfo {title} {{Excitation of molecular rotation by
  periodic microwave pulses. A testing ground for Anderson localization}},\
  }\href {https://doi.org/10.1063/1.450330} {\bibfield  {journal} {\bibinfo
  {journal} {Journal of Chemical Physics}\ }\textbf {\bibinfo {volume} {84}},\
  \bibinfo {pages} {2604} (\bibinfo {year} {1986})}\BibitemShut {NoStop}%
\bibitem [{\citenamefont {Ni}\ \emph {et~al.}(2017)\citenamefont {Ni},
  \citenamefont {Dadras}, \citenamefont {Lam}, \citenamefont {Shrestha},
  \citenamefont {Sadgrove}, \citenamefont {Wimberger},\ and\ \citenamefont
  {Summy}}]{Ni2017}%
  \BibitemOpen
  \bibfield  {author} {\bibinfo {author} {\bibfnamefont {J.}~\bibnamefont
  {Ni}}, \bibinfo {author} {\bibfnamefont {S.}~\bibnamefont {Dadras}}, \bibinfo
  {author} {\bibfnamefont {W.~K.}\ \bibnamefont {Lam}}, \bibinfo {author}
  {\bibfnamefont {R.~K.}\ \bibnamefont {Shrestha}}, \bibinfo {author}
  {\bibfnamefont {M.}~\bibnamefont {Sadgrove}}, \bibinfo {author}
  {\bibfnamefont {S.}~\bibnamefont {Wimberger}},\ and\ \bibinfo {author}
  {\bibfnamefont {G.~S.}\ \bibnamefont {Summy}},\ }\bibfield  {title} {\bibinfo
  {title} {Hamiltonian ratchets with ultra-cold atoms},\ }\href
  {https://doi.org/https://doi.org/10.1002/andp.201600335} {\bibfield
  {journal} {\bibinfo  {journal} {Annalen der Physik}\ }\textbf {\bibinfo
  {volume} {529}},\ \bibinfo {pages} {1600335} (\bibinfo {year}
  {2017})}\BibitemShut {NoStop}%
\bibitem [{\citenamefont {Sadgrove}\ \emph {et~al.}(2007)\citenamefont
  {Sadgrove}, \citenamefont {Horikoshi}, \citenamefont {Sekimura},\ and\
  \citenamefont {Nakagawa}}]{Sadgrove07}%
  \BibitemOpen
  \bibfield  {author} {\bibinfo {author} {\bibfnamefont {M.}~\bibnamefont
  {Sadgrove}}, \bibinfo {author} {\bibfnamefont {M.}~\bibnamefont {Horikoshi}},
  \bibinfo {author} {\bibfnamefont {T.}~\bibnamefont {Sekimura}},\ and\
  \bibinfo {author} {\bibfnamefont {K.}~\bibnamefont {Nakagawa}},\ }\bibfield
  {title} {\bibinfo {title} {Rectified momentum transport for a kicked
  bose-einstein condensate},\ }\href
  {https://doi.org/10.1103/PhysRevLett.99.043002} {\bibfield  {journal}
  {\bibinfo  {journal} {Phys. Rev. Lett.}\ }\textbf {\bibinfo {volume} {99}},\
  \bibinfo {pages} {043002} (\bibinfo {year} {2007})}\BibitemShut {NoStop}%
\bibitem [{\citenamefont {Sadgrove}\ \emph {et~al.}(2008)\citenamefont
  {Sadgrove}, \citenamefont {Kumar},\ and\ \citenamefont
  {Nakagawa}}]{PhysRevLett.101.180502}%
  \BibitemOpen
  \bibfield  {author} {\bibinfo {author} {\bibfnamefont {M.}~\bibnamefont
  {Sadgrove}}, \bibinfo {author} {\bibfnamefont {S.}~\bibnamefont {Kumar}},\
  and\ \bibinfo {author} {\bibfnamefont {K.}~\bibnamefont {Nakagawa}},\
  }\bibfield  {title} {\bibinfo {title} {Enhanced factoring with a
  {Bose-Einstein Condensate}},\ }\href
  {https://doi.org/10.1103/PhysRevLett.101.180502} {\bibfield  {journal}
  {\bibinfo  {journal} {Phys. Rev. Lett.}\ }\textbf {\bibinfo {volume} {101}},\
  \bibinfo {pages} {180502} (\bibinfo {year} {2008})}\BibitemShut {NoStop}%
\bibitem [{\citenamefont {Oskay}\ \emph {et~al.}(2003)\citenamefont {Oskay},
  \citenamefont {Steck},\ and\ \citenamefont {Raizen}}]{OSKAY2003409}%
  \BibitemOpen
  \bibfield  {author} {\bibinfo {author} {\bibfnamefont {W.~H.}\ \bibnamefont
  {Oskay}}, \bibinfo {author} {\bibfnamefont {D.~A.}\ \bibnamefont {Steck}},\
  and\ \bibinfo {author} {\bibfnamefont {M.~G.}\ \bibnamefont {Raizen}},\
  }\bibfield  {title} {\bibinfo {title} {Timing noise effects on dynamical
  localization},\ }\href
  {https://doi.org/https://doi.org/10.1016/S0960-0779(02)00302-8} {\bibfield
  {journal} {\bibinfo  {journal} {Chaos, Solitons and Fractals}\ }\textbf
  {\bibinfo {volume} {16}},\ \bibinfo {pages} {409} (\bibinfo {year}
  {2003})}\BibitemShut {NoStop}%
\bibitem [{\citenamefont {d'Arcy}\ \emph {et~al.}(2001)\citenamefont {d'Arcy},
  \citenamefont {Godun}, \citenamefont {Oberthaler}, \citenamefont
  {Cassettari},\ and\ \citenamefont {Summy}}]{Oxford2001}%
  \BibitemOpen
  \bibfield  {author} {\bibinfo {author} {\bibfnamefont {M.~B.}\ \bibnamefont
  {d'Arcy}}, \bibinfo {author} {\bibfnamefont {R.~M.}\ \bibnamefont {Godun}},
  \bibinfo {author} {\bibfnamefont {M.~K.}\ \bibnamefont {Oberthaler}},
  \bibinfo {author} {\bibfnamefont {D.}~\bibnamefont {Cassettari}},\ and\
  \bibinfo {author} {\bibfnamefont {G.~S.}\ \bibnamefont {Summy}},\ }\bibfield
  {title} {\bibinfo {title} {Quantum enhancement of momentum diffusion in the
  delta-kicked rotor},\ }\href {https://doi.org/10.1103/PhysRevLett.87.074102}
  {\bibfield  {journal} {\bibinfo  {journal} {Phys. Rev. Lett.}\ }\textbf
  {\bibinfo {volume} {87}},\ \bibinfo {pages} {074102} (\bibinfo {year}
  {2001})}\BibitemShut {NoStop}%
\end{thebibliography}


%

\end{document}